\documentstyle[12pt]{article}
\newcommand{\mysection}[1]{\section{#1}
            \setcounter{equation}{0}\setcounter{figure}{0}}

\def\5{\bar }  \def\6{\partial } \def\7{\tilde }

\begin{document}


\begin{titlepage}

\vfill

\begin{flushright}
UB--ECM--PF--97/28\\
hep-th/9712125\\
to appear in Nucl. Phys. B
\end{flushright}

\vfill

\begin{center}
{\LARGE
Cohomological analysis of bosonic D-strings and 2d sigma models coupled
to abelian gauge fields
}
\end{center}

\begin{center}
{\large
Friedemann Brandt, Joaquim Gomis, Joan Sim\'on
}
\end{center}

\begin{center}
{\sl
Departament ECM,
Facultat de F\'{\i}sica,
Universitat de Barcelona and 
Institut de F\'{\i}sica d'Altes Energies,
Diagonal 647,
E-08028 Barcelona, Spain.\\
E-mail: brandt@itp.uni-hannover.de, gomis@ecm.ub.es, jsimon@ecm.ub.es.
}
\end{center}

\begin{abstract}
We analyse completely the BRST cohomology on local functionals for 
two dimensional sigma models coupled to abelian world sheet gauge fields, 
including effective bosonic D-string models described by Born-Infeld actions. 
In particular we prove that the rigid symmetries of such models are 
exhausted by the solutions to generalized Killing vector equations 
which we have presented recently, and provide all the consistent first 
order deformations and candidate gauge anomalies of the models under study.
For appropriate target space geometries we find nontrivial
deformations both of the abelian gauge transformations and of the world sheet 
diffeomorphisms, and antifield dependent candidate anomalies for both types 
of symmetries separately, as well as mixed ones. 
\end{abstract}

\vfill


\vfill

\end{titlepage}


\mysection{Introduction}\label{intro}

D-branes \cite{joe} play a crucial role in 
recent advances towards understanding nonperturbative properties
of string and supersymmetric quantum field theories. 
The low energy effective action of 
a D-brane involves a Born--Infeld type action containing 
the coordinates of the brane and a $U(1)$
world-volume gauge field. In the case of IIB D-branes, it has been recently
suggested that, for
a manifestly $SL(2,Z)$ invariant formulation, the action should
contain a world-volume field for every background
gauge potential \cite{paul,cederwall}.
In the case of the D-string one then should introduce two $U(1)$ gauge 
fields on the world-sheet.

Inspired by these developments, we consider in this paper
a class of models including
(effective) bosonic D-strings described by
Born--Infeld type actions, but not restricted to them.
Rather, the models under study are characterized only
by their field content,
gauge symmetries and a locality requirement. 
Specifically, we consider models described in terms 
of a set of world-sheet 
scalar fields $X^M$ (``target space coordinates''), a set of world-sheet
gauge fields $A_\mu^a$, and an auxiliary world-sheet metric
$\gamma_{\mu\nu}$. The action functionals for such models
are required to be local, i.e.
polynomial in derivatives of the fields, and
invariant under world-sheet diffeomorphisms,
local Weyl transformations of $\gamma_{\mu\nu}$ and
abelian gauge transformations of the $A_\mu^a$ (the local Weyl
symmetry is spurious in the sense that it can be avoided if 
one works with two ``Beltrami variables''
$h_{++}$, $h_{--}$ instead of the three components of 
$\gamma_{\mu\nu}$, see section \ref{contractions}).
The classification of all action functionals satisfying these
requirements will be part of our results. This includes
Born--Infeld type actions,
as they can be written in a form which is local
in the above sense by means of scalar auxiliary fields counting
among the $X^M$, see section \ref{action}.

The main purpose of the paper is to investigate these models with regard 
to their symmetries. Specifically, we aim at
a classification of their rigid (= global) symmetries and dynamical
conservation laws, and
of the possible anomalies and nontrivial continuous deformations
of their gauge (= local) symmetries. Studying such deformations covers
the determination of possible gauge invariant 
interactions between world sheet gauge fields and the target fields $X^M$, 
and has useful applications within the renormalization program for 
effective gauge theories \cite{stevequim}. 

These issues are analysed systematically by
computing the BRST cohomology in the space of local functionals
of the fields and their antifields. The rigid symmetries
and dynamical conservation laws arise from this cohomology
at negative ghost numbers \cite{bbh1}, the action functionals
and consistent (first order) deformations at ghost number 0 
\cite{bh}, and the candidate anomalies at ghost number 1 \cite{anos}.
In fact we will compute the cohomology at all ghost numbers and
discuss afterwards the results in detail for ghost numbers $\leq 1$.

We have organized the paper as follows.
In section \ref{setup} we describe the models and the cohomological
problem to be studied.
In section \ref{contractions} we perform the first part of 
the cohomological analysis
for which the explicit knowledge of a classical action is not needed.
The action functionals themselves 
are then determined in section \ref{action}.
We then complete the cohomological analysis and summarize
the result in a compact form for all ghost numbers
in section \ref{Result} (the computation itself
is relegated to an appendix). In section \ref{discussion}
we spell out explicitly the physically most important solutions 
(those with ghost numbers $\leq 1$) and comment on them.
To illustrate these results we discuss in section \ref{example}
a class of models with only one abelian
gauge field. The paper is ended by
some conclusions in section \ref{conclusions}, and
three technical appendices.

\mysection{The cohomological problem}
\label{setup}

As mentioned already, the field content of the models we are going 
to study is
given by the world-sheet metric $\gamma_{\mu\nu}$ ($\mu=+,-$),
a set of scalar fields $X^M$ -- some of which may be
auxiliary fields -- and a set of 
abelian gauge fields
$A^a_\mu$. The number of scalar and gauge fields
is not specified, i.e. our approach covers models
with any number of such fields. We impose
invariance under world-sheet diffeomorphisms, local
Weyl transformations of the world-sheet metric, and
abelian gauge transformations of the
$A^a_\mu$. The ghost fields associated with these
gauge symmetries are denoted by $\xi^\mu$ (diffeomorphism ghosts),
$c$ (Weyl ghost) and $C^a$ (abelian ghosts) respectively.
This fixes the field content to
\[
\{\Phi^A\}=\{\gamma_{\mu\nu}\, ,\, X^M\, ,\, A^a_\mu\, ,\, 
\xi^\mu\, ,\, c\, ,\, C^a\}.
\]
The BRST transformations of these fields corresponding to the
above mentioned gauge symmetries are
\begin{eqnarray}
s\gamma_{\mu\nu}&=&\xi^\rho\6_\rho \gamma_{\mu\nu}
+\6_\mu\xi^\rho\cdot \gamma_{\rho\nu}
+\6_\nu\xi^\rho\cdot \gamma_{\mu\rho}+c\, \gamma_{\mu\nu}
\nonumber\\
sX^M &=&\xi^\mu\6_\mu X^M
\nonumber\\
sA^a_\mu&=& \xi^\nu\6_\nu A^a_\mu 
+\6_\mu\xi^\nu\cdot A^a_\nu+\6_\mu C^a
\nonumber\\
s\xi^\mu &=&\xi^\nu\6_\nu\xi^\mu
\nonumber\\
sc&=& \xi^\mu\6_\mu c
\nonumber\\
sC^a&=& \xi^\mu\6_\mu C^a.
\label{brs1}\end{eqnarray}

According to the standard prescription
of the field-antifield formalism \cite{bv,henteit,report}, the 
BRST transformations of the antifields are given by
\begin{equation}
s\Phi^*_A\equiv\left({\cal S},\Phi^*_A\right)=
\frac {\delta^R{\cal S}}{\delta \Phi^A}
\label{brs4}\end{equation}
where ${\cal S}$ is the proper solution of 
the (classical) master equation. As the gauge algebra is closed,
${\cal S}$ reads in our case simply
\begin{equation}
{\cal S}=\int d^2\sigma\, \left[L_0-(s\Phi^A)\Phi^*_A\right]
\label{S}\end{equation}
where $L_0$ is the integrand of a classical action
with (only) the above gauge symmetries.
For later purposes we decompose the BRST-transformations
into parts with different so-called antighost numbers.
In this case the decomposition has only two pieces, denoted
by $\delta$ and $\gamma$ respectively where $\delta$ is the
field theoretical Koszul--Tate differential \cite{KT} which acts
nontrivially only on the antifields,
\begin{equation}
s=\delta + \gamma ,\quad 
s\Phi^A=\gamma\Phi^A,\quad s\Phi^*_A=(\delta+\gamma)\Phi^*_A \ .
\end{equation}
Only $\delta$ depends on the explicit form of $ L_0$,
\begin{eqnarray}
\delta \phi^*_i &=& \frac{\hat \6L_0}{\hat \6 \phi^i}\quad
\mbox{for}\quad \{\phi^i\}=\{X^M\, ,\, A_\mu^a\, ,\, \gamma_{\mu\nu}\}
\nonumber\\ 
\delta \xi^*_\mu &=& X^*_M \6_\mu X^M
+\gamma^{\rho\sigma *} \6_\mu \gamma_{\rho\sigma}
+A^{\nu *}_a \6_\mu A_\nu^a
+\6_\nu (2\gamma_{\mu\rho} \gamma^{\nu\rho *}+A_\mu^a A^{\nu *}_a)
\nonumber\\
\delta c^* &=& \gamma_{\mu\nu}\gamma^{\mu\nu *}
\nonumber\\
\delta C^*_a &=& -\6_\mu A^{\mu *}_a
\label{delta}\end{eqnarray}
where ${\hat \6L_0}/{\hat \6 \phi^i}$ is the Euler--Lagrange
derivative of $L_0$ with respect to $\phi^i$.
In contrast, $\gamma$ is determined already by the 
gauge symmetries only,
\begin{eqnarray}
\gamma X^*_M &=&\6_\mu(\xi^\mu X^*_M)
\nonumber\\
\gamma A^{\mu *}_a &=&
\6_\nu(\xi^\nu A^{\mu *}_a)
-\6_\nu\xi^\mu\cdot A^{\nu *}_a
\nonumber\\
\gamma \gamma^{\mu\nu *} &=&
\6_\rho(\xi^\rho  \gamma^{\mu\nu *})
-2\6_\rho\xi^{(\nu} \cdot \gamma^{\mu)\rho *}-c\, \gamma^{\mu\nu *}
\nonumber\\
\gamma \xi^*_\mu &=& \6_\nu(\xi^\nu \xi^*_\mu)
+\xi^*_\nu \6_\mu\xi^\nu+C^*_a \6_\mu C^a+c^* \6_\mu c
\nonumber\\
\gamma c^* &=& \6_\mu(\xi^\mu c^*)
\nonumber\\
\gamma C^*_a &=& \6_\mu(\xi^\mu C^*_a).
\label{gamma}\end{eqnarray}

The cohomological problem to be studied is defined on
local functionals. These are local functions integrated over
the world-sheet,
\begin{equation}
W^g=\int d^2\sigma\, w,\quad \mbox{gh}(w)=g
\label{functional}
\end{equation}
where $g$ denotes the ghost number ($\mbox{gh}$) and
the integrands are required to depend
polynomially on derivatives of all the fields and antifields.
A restriction on the number or order of derivatives
that may occur is however {\em not} imposed. The dependence
on the undifferentiated fields
$X^M$ and $\gamma_{\mu\nu}$ may of course be nonpolynomial.
By definition,
a local functional (\ref{functional}) is called BRST invariant if
the BRST transformation of its integrand is
a total derivative,
\begin{equation}
s\, w=\6_\mu v^\mu,
\label{i1}
\end{equation}
where the $v^\mu$ are required to be local functions too. Solutions
of the form $s\hat w+\6_\mu \hat v^\mu$
are called trivial whenever $\hat w$ and $\hat v^\mu$ are
local functions, and two solutions are called equivalent ($\simeq$) 
if they differ by a trivial solution,
\begin{equation}
w \simeq w'\quad:\Leftrightarrow\quad w-w'=s\,\hat w+\6_\mu \hat v^\mu.
\label{i2}
\end{equation}

\mysection{Contracting homotopies and conformal structure}
\label{contractions}

The first part of the cohomological analysis consists in the
construction of contracting homotopies which reduce the
cohomological problem considerably and do not depend on the
specific choice of the classical action (only the knowledge
of the gauge transformations is needed). It applies a
technique described in \cite{ten} which reveals 
in this case a `conformal structure'
of the models under study and reduces the problem to the
cohomology in a space of conformal
tensor fields and generalized connections.
This conformal structure enables us to reduce the
analysis further to a ``conformally invariant'' cohomological problem
in a manner that parallels
the analysis of sigma models performed in \cite{paper1}.
We shall not repeat in detail the arguments and constructions
used in \cite{ten} and \cite{paper1};
rather we shall 
focus on the basic ideas and describe afterwards their outcome
for the models studied here.
These ideas are:
\begin{enumerate}
\item 
The formulation of the cohomological problem in the
jet bundle of the fields and antifields. Essentially this
means that the fields, antifields and all their
derivatives are considered as local coordinates
of an infinite jet space $J^\infty$.
\item 
The mapping of the BRST cohomology in the space of local functionals
to the cohomology of 
\begin{equation} \7s=s+d \end{equation}
in the space of local total forms on $J^\infty$
(``$\7s$-cohomology'' for short). 
A local total form is simply a sum of local forms
with different form degrees and ghost numbers.
\item
The construction of a complete set of new local jet coordinates
$\{{\cal U}^\ell,{\cal V}^\ell,{\cal W}^i\}$ satisfying
\begin{equation}
\7s\, {\cal U}^\ell={\cal V}^\ell,\quad \7s\, {\cal W}^i={\cal R}^i({\cal W}).
\label{c1}
\end{equation}
This allows to contract the $\7s$-cohomology on local total forms
(locally) to the $\7s$-cohomology on local total forms constructed
solely of the ${\cal W}$'s. 
\item
A second very powerful contracting homotopy
which eliminates also almost all of the ${\cal W}$'s and
reduces the cohomological problem to the $\7s$-cohomology
in the space of local total forms with vanishing ``conformal weights'' 
constructed of the few remaining ${\cal W}$'s.
\end{enumerate}

The mapping to the $\7s$-cohomology  just compresses the
analysis of the so-called descent equations following from
(\ref{i1}). Namely, written in terms of differential
forms, (\ref{i1}) reads $s\omega_2+d\omega_1=0$ where
$\omega_2=d\sigma^+d\sigma^- w$, $\omega_1=
d\sigma^\mu \epsilon_{\nu\mu}v^\nu$
(with $\epsilon_{-+}=1$). By the standard arguments one
concludes then the existence of a 0-form $\omega_0$ such that
altogether
\begin{equation}
s\,\omega_2+d\,\omega_1=0,\quad s\,\omega_1+d\,\omega_0=0,\quad
s\,\omega_0=0.
\label{descent}
\end{equation}
These are the descent equations. They are equivalent to
\begin{equation}
\7s\omega=0,\quad \omega=\omega_2+\omega_1+\omega_0\ .
\label{OMEGA}
\end{equation}
When switching from $s$ to $\7s$, it is natural to
introduce a {\em total degree} ($\mbox{totdeg}$) which is
the sum of the ghost number and the form degree
(i.e.\ $\7s$ has total degree 1),
\begin{equation}
\mbox{totdeg} = \mbox{gh} + \mbox{formdeg}\ .
\end{equation}
Hence, the integrands
of the BRST invariant local functionals with ghost number $g$ are just the
2-forms contained in the corresponding $\7s$-invariant
local total forms with total degree $g+2$. 

The crucial properties of jet coordinates satisfying (\ref{c1})
are of course
that the ${\cal U}$'s and ${\cal V}$'s form ``$\7s$-doublets'', and
that $\7s$ leaves the space of the ${\cal W}$'s invariant.
The construction of these jet coordinates given below
is in fact technically
the most involved part of our analysis. 
The differences to an analogous
construction in \cite{paper1} are 
(a) the presence of the abelian gauge and ghost fields
and their antifields which amounts to the finding of
${\cal U}$'s, ${\cal V}$'s and ${\cal W}$'s corresponding to these extra
fields, and (b) the switch from $s$ to $\7s$.
(b) is actually straightforward as
the $s$-transformations turn into $\7s$-transformations
by the replacement $\xi^\mu \rightarrow \xi^\mu+d\sigma^\mu$. 
Nevertheless, switching from $s$ to $\7s$ simplifies some arguments, 
such as the proof that
explicit world-sheet coordinates can be removed from all the
cocycles (as they count among the ${\cal U}$'s).

In order to construct the ${\cal U}$'s, ${\cal V}$'s and ${\cal W}$'s
we first replace the undifferentiated components of the
world-sheet metric and of the ghosts by the following
new jet coordinates,
\begin{eqnarray}
h_{\pm\pm}&=&\gamma_{\pm\pm}(\gamma_{+-}+\sqrt{\gamma})^{-1}
\nonumber\\
e&=&\sqrt{\gamma}
\nonumber\\
\7\xi^\pm&=&\xi^\pm+d\sigma^\pm +h_{\mp\mp}(\xi^\mp+d\sigma^\mp)
\nonumber\\
\7c&=&c\, e+\6_\mu (\xi^\mu e)+d\sigma^\mu \6_\mu e
\nonumber\\
\7C^a&=&C^a+(\xi^\mu+d\sigma^\mu) A^a_\mu\ .
\label{red1}\end{eqnarray}
The $h_{\pm\pm}$ may be viewed as ``Beltrami variables''. 
The $\7s$-transformations of the fields read in the new basis:
\begin{eqnarray}
\7s\, h_{\pm\pm}&=&(\6_\pm-h_{\pm\pm}\6_\mp
+\6_\mp h_{\pm\pm}\cdot)\,\7\xi^\mp
\nonumber\\
\7s\, e&=&\7c
\nonumber\\
\7s\, A^a_\pm &=&\6_\pm \7C^a\mp(\7\xi^\mp-h_{\pm\pm}\7\xi^\pm)F^a
\nonumber\\
\7s\, \7\xi^\pm&=&\7\xi^\pm\6_\pm \7\xi^\pm
\nonumber\\
\7s\, \7c&=&0
\nonumber\\
\7s\, \7C^a&=&\7\xi^+\7\xi^- F^a
\nonumber\\
\7s\, X^M&=& \7\xi^\mu {\cal D}_\mu X^M
\label{c2}\end{eqnarray}
where ${\cal D}_\mu$ are covariant derivatives given explicitly below, and 
\begin{equation}
F^a=\frac 12\, (1-h_{++}h_{--})^{-1}\,\epsilon^{\mu\nu}F^a_{\mu\nu}\, ,
\quad F^a_{\mu\nu}=\6_\mu A^a_\nu-\6_\nu A^a_\mu\ .
\label{F}\end{equation}
Next one constructs
new jet coordinates replacing the derivatives of the fields.
Taking also the world-sheet coordinates $\sigma^\mu$
and differentials $d\sigma^\mu$ into account (one has
$\7s\sigma^\mu=d\sigma^\mu$), this yields the
new jet coordinates with nonnegative total degrees,
\begin{eqnarray}
\{{\cal U}_0^\ell\}&=&\{\sigma^\mu,\,\6_+^m\6_-^n h_{\pm\pm},\,
\6_+^m\6_-^n e,\,
\6_+^m\6_-^n A^a_-,\, \6_+^m A^a_+:\ m,n=0,1,\dots\}
\nonumber\\
\{{\cal V}_1^\ell\}&=&\{\7s\, {\cal U}_0^\ell\}
\nonumber\\
\{{\cal W}_1^i\}&=&\{\7C^a,\, \6_\pm^m\7\xi^\pm : 
m=0,1,\dots\}
\nonumber\\
\{{\cal W}_0^i\}&=&\{ {\cal D}_+^m{\cal D}_-^n X^M,\, 
{\cal D}_+^m{\cal D}_-^n F^a:\
m,n=0,1,\dots\}
\label{basis1}
\end{eqnarray}
where subsripts 0 or 1 indicate the total degree, and ${\cal D}_\pm$
are commuting covariant derivatives
\begin{equation}
{\cal D}_\pm=(1-h_{++}h_{--})^{-1}\left[
\6_\pm-h_{\pm\pm}\6_\mp-\sum_{m\geq 0}(H_\mp^m L^\mp_m
-h_{\pm\pm}H_\pm^m L^\pm_m) \right].
\label{covderiv}\end{equation}
Here
\[
H_\pm^m=\frac 1{(m+1)!}\, \6_\pm^{m+1}h_{\mp\mp}
\]
and
\begin{eqnarray}
L^\pm_r ({\cal D}_\pm^m{\cal D}_\mp^n X^M) &=&
\left\{
\begin{array}{cl}
\frac{m!}{(m-r-1)!}\, {\cal D}_\pm^{m-r}{\cal D}_\mp^n X^M & 
       \mbox{for } 0\leq r<m
\\
0 & \mbox{for } r\geq m
\end{array}\right.
\nonumber\\
L^\pm_r ({\cal D}_\pm^m{\cal D}_\mp^n F^a) &=&
\left\{
\begin{array}{cl}
\frac{(m+1)!}{(m-r)!}\, {\cal D}_\pm^{m-r}{\cal D}_\mp^n F^a & 
\mbox{for } 0\leq r\leq m
\\
0 & \mbox{for } r> m.
\end{array}\right.
\label{vir1}
\end{eqnarray}
The explicit form of
the ${\cal W}_0^i$ in terms of the fields and their derivatives
is obtained from
(\ref{covderiv}) and (\ref{vir1}) in a recursive manner. This is
possible because the
right hand sides of (\ref{vir1}) do not contain powers of ${\cal D}_\pm$
exceeding $m$. For instance one gets
\begin{eqnarray}
{\cal D}_+ X^M&=&(1-h_{++}h_{--})^{-1}(\6_+ -h_{++}\6_-)X^M
\nonumber\\
{\cal D}_- X^M&=&(1-h_{++}h_{--})^{-1}(\6_- -h_{--}\6_+)X^M
\nonumber\\
{\cal D}_+{\cal D}_-X^M&=&(1-h_{++}h_{--})^{-1}(\6_+ -\6_-h_{++}){\cal D}_-X^M
\label{Ts}\end{eqnarray}
where in the last line $\6_-$ acts both on $h_{++}$ and ${\cal D}_-X^M$.
It is also easy to verify by induction that all ${\cal W}_0^i$ constructed in
this way are local functions because
only finitely many summands of the infinite
sums in (\ref{covderiv}) are actually nonvanishing thanks to (\ref{vir1}).
We will show below that one can choose the representatives
of the cohomology classes such that the only ${\cal W}_0^i$
that contribute to them are
$X^M$, $F^a$ and the covariant derivatives of $X^M$ listed in
(\ref{Ts}). Hence, the explicit form of all other ${\cal W}_0^i$ 
will actually not matter later on.
Nevertheless the fact that they exist as
local functions of the fields is crucial for our arguments.

The ${\cal W}^i_0$ can be viewed as conformal tensor fields,
and the ${\cal D}_\mu$ as conformal covariant derivatives.
Indeed, using (\ref{vir1}) and the identification
\begin{equation}
L^\mu_{-1}\equiv{\cal D}_\mu
\label{vir2}\end{equation} 
one easily
checks that the $L$'s represent on the ${\cal W}^i_0$
two copies of
the regular part of the Wit algebra,
\begin{eqnarray}
\mbox{\bf [}\, L^\pm_m\, ,\, L^\pm_n\, 
\mbox{\bf ]}=(m-n)L^\pm_{m+n}\ ,\quad
\mbox{\bf [}\, L^+_m\, ,\, L^-_n\, \mbox{\bf ]}=0\quad 
(m,n=-1,0,1,\dots).
\label{vir3}\end{eqnarray}
The occurrence of this infinite dimensional
algebra reflects the infinite dimension of the
conformal group in two dimension and
is related to the Weyl invariance which removes
the `conformal' degree of freedom contained in $\gamma_{\mu\nu}$.
Indeed, in the above approach this degree of freedom is associated
with the variable $e=\sqrt \gamma$. Together with the Weyl ghost 
field, $e$ drops completely out of
the cohomological analysis, as these fields and all their
derivatives form $\7s$-doublets. 
In fact, we could have worked from the beginning
with the `Beltrami' fields $h_{++}$ and $h_{--}$
instead of using the three components of
$\gamma_{\mu\nu}$ and imposing Weyl invariance. 
Now, (\ref{c2}) suggests to view $h_{++}$ and $h_{--}$
as gauge fields for `chiral diffeomorphisms', as 
$\7s h_{++}$ contains $\6_+\7\xi^-$ while
$\7s h_{--}$ contains $\6_-\7\xi^+$, but there are 
no `gauge fields' corresponding to $\6_+\7\xi^+$ 
and $\6_-\7\xi^-$. As a consequence, the $m$th order
derivatives of $h_{++}$ and $h_{--}$ pair off
with the $(m+1)$th order derivatives of the
$\xi^\mu$ except for $\6_+^{m+1}\7\xi^+$ and $\6_-^{m+1}\7\xi^-$.
The undifferentiated ghost fields $\7\xi^\pm$
and their non-paired derivatives $\6_\pm^{m+1}\7\xi^\pm$
correspond to the $L^\pm_m$.

{}From (\ref{c2}) it is obvious  
that the ${\cal W}^i_1$ satisfy (\ref{c1}). The
${\cal W}^i_0$ satisfy (\ref{c1}) too as they transform according to
\begin{equation}
\7s\, {\cal W}^i_0=\sum_{m\geq -1} \frac 1{(m+1)!} \left(
\6_+^{m+1}\7\xi^+\cdot L^+_m
+\6_-^{m+1}\7\xi^-\cdot L^-_m\right)\, {\cal W}^i_0\ .
\label{sT}\end{equation}
Again, these expressions are local as only finitely many summands 
are nonvanishing thanks to (\ref{vir1}). 

Next we construct appropriate new local jet coordinates replacing
one by one the antifields and all their derivatives such that
(\ref{c1}) holds. For this construction the
knowledge of the classical action is not needed, as
$\7\gamma {\cal W}^i=R^i({\cal W})$
(with $\7\gamma=\gamma+d$) implies $\7s {\cal W}^i={\cal R}^i({\cal W})$ 
in our case
as a consequence of $\7s=\delta + \7\gamma$ and
$\delta\7\gamma+\7\gamma\delta=0$. Furthermore
$\7\gamma$ does not involve the
explicit form of the classical action but only the
gauge symmetries, cf. section \ref{setup}. This is actually the reason that
one can use the same ${\cal U}$'s, ${\cal V}$'s and ${\cal W}$'s 
for all actions
with the field content and gauge symmetries that we have imposed.

The construction of new jet coordinates
replacing the antifields $\gamma^{\mu\nu *}$, $\xi^*_\mu$,
$c^*$ and all their derivatives
has been given in \cite{paper1}. 
Their explicit form will not be needed later on as none of them will
contribute nontrivially to the cohomology.
We will denote them
by $\6_+^m\6^n_-e^*$, $\6_+^m\6^n_-\7c^*$, $h^{\pm\pm *}_{m,n}$,
$\xi^*_{\mu m,n}$ ($m,n=0,1,\dots$). The $\6_+^m\6^n_-\7c^*$ and
$\6_+^m\6^n_-e^*$ count among the ${\cal U}$'s and ${\cal V}$'s
respectively (due to $\7s\7c^*=e^*$), whereas the $h^{\pm\pm *}_{m,n}$
and $\xi^*_{\mu m,n}$ are ${\cal W}$'s. The new jet coordinates
replacing the undifferentiated antifields 
$X^*_M$, $A^{\mu *}_a$ and $C^*_a$ are
\begin{eqnarray} 
\hat X^*_M &=&(1-h_{++}h_{--})^{-1}X^*_M
\nonumber\\
\hat A^{\pm*}_a&=&(1-h_{++}h_{--})^{-1}(A^{\pm*}_a+h_{\mp\mp}A^{\mp*}_a)
\nonumber\\
\hat C^*_a&=&(1-h_{++}h_{--})^{-1}C^*_a\ .
\label{basis3}\end{eqnarray}
The partial derivatives of $X^*_M$, $A^{\mu *}_a$ and $C^*_a$ 
are replaced by covariant derivatives of the redefined antifields
(\ref{basis3}). These covariant derivatives are constructed
recursively in the same manner as those of $X^M$ and
$F^a$ above, where $L_r^\pm$ acts on the covariant derivatives 
of $\hat X^*_M$, $\hat A^{\mp *}_a$ and $\hat C^*_a$ 
as on those of $F^a$ 
in (\ref{vir1}), while it acts on the covariant derivatives
of $\hat A^{\pm *}_a$ as on those of $X^M$.
In particular this yields
\begin{equation}
{\cal D}_\pm\hat A^{\pm*}_a=(1-h_{++}h_{--})^{-1}
(\6_\pm -\6_\mp h_{\pm\pm})\,\hat A^{\pm*}_a\ .
\label{basis4}
\end{equation}
Again, the explicit expressions for other covariant
derivatives of $\hat X^*_M$, $\hat C^*_a$ or $\hat A^{\mu *}_a$
will not be needed later on
as they will contribute only trivially to the cohomology.

$\hat X^*_M$, $\hat C^*_a$, $\hat A^{\mu *}_a$ and their covariant
derivatives count among the ${\cal W}$'s and we thus get the following
set of new jet coordinates replacing the antifields and all their
derivatives:
\begin{eqnarray}
\{{\cal U}^\ell_{-2}\}&=&\{ \6_+^m\6_-^n\7c^*:\  m,n=0,1,\dots\}
\nonumber\\
\{{\cal V}^\ell_{-1}\}&=&\{\6_+^m\6_-^ne^*:\  m,n=0,1,\dots\}
\nonumber\\
\{{\cal W}^i_{-1}\}&=&\{h^{\pm\pm *}_{m,n}\, ,\, 
{\cal D}_+^m{\cal D}_-^n\hat X^*_M\, ,\, 
{\cal D}_+^m{\cal D}_-^n\hat A^{\mu *}_a:\ m,n=0,1,\dots\}
\nonumber\\
\{{\cal W}^i_{-2}\}&=&\{ \xi^*_{\pm m,n}\, ,\,  
{\cal D}_+^m{\cal D}_-^n\hat C^*_a:\ m,n=0,1,\dots\}.
\end{eqnarray}
As mentioned above, the ${\cal W}^i_{-1}$ and ${\cal W}^i_{-2}$ 
satisfy (\ref{c1}) as a consquence of their 
$\7\gamma$-transformations.
The $\7\gamma$-transformations of
$\hat X^*_M$, $\hat A^{\mu *}_a$, $\hat C^*_a$ and their
covariant derivatives take the same form as the 
$\7s$-transformations of the ${\cal W}^i_{0}$ given in (\ref{sT}).

This completes the construction of the ${\cal U}$'s, ${\cal V}$'s 
and ${\cal W}$'s. As mentioned in the beginning, the
construction performed so far allows us now to eliminate all the
${\cal U}$'s and ${\cal V}$'s from the cohomology (at least locally) and
we are left with the ${\cal W}$'s. Next we show that actually only
very few ${\cal W}$'s contribute nontrivially to the cohomology.
To that end we extend the definition of 
$L_0^+$ and $L_0^-$ to all ${\cal W}$'s by
\begin{equation}
L_0^\pm=\left\{\7s\, ,\, \frac{\6}{\6(\6_\pm\7\xi^\pm)}\,\right\}.
\label{L0}\end{equation}
This reproduces the action of $L_0^\pm$ in (\ref{vir1}) and 
extends it such that all ${\cal W}$'s 
are eigenfunctions of $L_0^+$ and $L_0^-$. Indeed one has
\begin{equation}
L_0^\pm {\cal W}^i=w^{\pm}(i)\, {\cal W}^i
\end{equation}
where the eigenvalues $w^\pm(i)$ are integer
``conformal weights'' listed in table 1.
\[
\begin{array}{|c|c|c||c|c|c|}
\multicolumn{3}{c}{\mbox{Fields}}&\multicolumn{3}{c}{\mbox{Antifields}}\\
\hline
\mbox{totdeg}& {\cal W}^i & w^+(i),w^-(i) &\mbox{totdeg}& 
         {\cal W}^i & w^+(i),w^-(i) 
\rule{0em}{3ex}\\
\hline
1  & \6_+^m\7\xi^+ & m-1,0      & -2 & \xi^*_{+ m,n} & m+2,0 
\rule{0em}{3ex}\\
   & \6_-^m\7\xi^- & 0,m-1      &    & \xi^*_{- m,n} & 0,m+2 
\rule{0em}{3ex}\\
   & \7C^a        & 0,0        &    &{\cal D}_+^m{\cal D}_-^n\hat C^*_a & 
   m+1,n+1
\rule{0em}{3ex}\\
\hline
0 & {\cal D}_+^m{\cal D}_-^n X^M & m,n    & -1 &
  {\cal D}_+^m{\cal D}_-^n\hat X^*_M & m+1,n+1
\rule{0em}{3ex}\\
  &                 &           &    & h^{++*}_{m,n} & 0,n+2 
\rule{0em}{3ex}\\
  &                 &           &    & h^{--*}_{m,n} & m+2,0
\rule{0em}{3ex}\\
  &{\cal D}_+^m{\cal D}_-^n F^a & m+1,n+1&    &
  {\cal D}_+^m{\cal D}_-^n\hat A^{+ *}_a& m,n+1
\rule{0em}{3ex}\\
  &                 &           &    &
  {\cal D}_+^m{\cal D}_-^n\hat A^{- *}_a& m+1,n   
\rule{0em}{3ex}\\
\hline
\multicolumn{6}{c}{}\\
\multicolumn{6}{c}{\mbox{Table 1: Total degrees and conformal 
weights of the ${\cal W}$'s}}
\end{array}
\]
As $L^+_0$ and $L^-_0$ are 
anticommutators with $\7s$, they establish contracting
homotopies for the
$\7s$-cohomology in the space of local total forms constructed
of the ${\cal W}$'s, i.e., this cohomology can be 
nontrivial only
in the intersection of the kernels of $L^+_0$ and $L^-_0$.
Hence, all nontrivial representatives of the $\7s$-cohomology
can be brought to the form $\omega({\cal W})$ where $\omega({\cal W})$
has vanishing conformal weights,
\begin{equation}
\7s\,\alpha({\cal W})=0\ \Rightarrow\ \alpha({\cal W})=
\omega({\cal W})+\7s\,\beta({\cal W}),
\ L_0^\pm\omega=0.
\label{0,0}\end{equation}
It is in fact easy to construct the most general local 
$\omega({\cal W})$ with vanishing conformal weights. Indeed,
table 1 shows that all ${\cal W}$'s have a nonnegative conformal weight
$w^+$ or $w^-$, except for
$\7\xi^+$ and $\7\xi^-$ which have weights
$(-1,0)$ and $(0,-1)$ respectively. $\7\xi^+$ and $\7\xi^-$
are thus the only ${\cal W}$'s which can compensate for positive 
conformal weights of other ${\cal W}$'s. As $\7\xi^+$ and $\7\xi^-$
are Grassmann odd, each of them can appear
at most once in a monomial of ${\cal W}$'s. Hence, 
a ${\cal W}^i$ with $w^+(i)>1$ or $w^-(i)>1$ cannot appear in
a {\em local} total form with vanishing conformal weights.
$\omega$ can thus only depend on the undifferentiated
$\7\xi^\mu$ and on those ${\cal W}$'s with conformal weights $(0,0)$,
$(1,0)$, $(0,1)$ or $(1,1)$ respectively. Furthermore
the ${\cal W}$'s with conformal weights $(1,0)$, $(0,1)$ or $(1,1)$
appear necessarily together
with $\7\xi^+$, $\7\xi^-$ and $\7\xi^+\7\xi^-$
respectively. We conclude that $\omega$ can be
constructed solely out of the following quantities:
\begin{eqnarray}
\{z^s\}&=&\{X^M\, ,\, \7\Theta_+\, ,\, \7\Theta_-\, ,\, \7C^a \}
\nonumber\\
\{y_\pm^A\}&=&\{\7X^M_\pm\, ,\, \7Y_\pm\, ,\, \7A^{\mp *}_{a \pm}\}
\nonumber\\
\{y_{+-}^\Delta\}&=&\{\7X^M_{+-}\, ,\, \7F^a_{+-}\, ,\, 
\7X^*_{M+-}\, ,\, \7A^{\mu *}_{a +-}\, ,\, \7C^*_{a +-}\}
\label{compact}
\end{eqnarray}
where
\begin{eqnarray}
&
\7\Theta_\pm=\6_\pm\7\xi^\pm,
&
\nonumber\\
&
\7X^M_\pm=\7\xi^\pm{\cal D}_\pm X^M,\quad 
\7Y_\pm=\7\xi^\pm\6_\pm^2\7\xi^\pm,\quad
\7A^{\mp *}_{a \pm} = \7\xi^\pm\hat A^{\mp *}_a\, ,
&
\nonumber\\
&
\7X^M_{+-}=\7\xi^+\hat \xi^-{\cal D}_+{\cal D}_-X^M,\quad
\7F^a_{+-}=\7\xi^+\7\xi^-F^a,
&
\nonumber\\
&
\7X^*_{M+-} = \7\xi^+\7\xi^-\hat X^*_M\, ,\quad
\7A^{\pm *}_{a +-} =  \7\xi^+\7\xi^-{\cal D}_\pm\hat A^{\pm *}_a\, ,\quad
\7C^*_{a +-} = \7\xi^+\7\xi^-\hat C^*_a\ .
&
\label{Q1}\end{eqnarray}
$\{z^s\}$ contains all ${\cal W}$'s with
conformal weights $(0,0)$, $\{y_+^A\}$ ($\{y_-^A\}$) contains
all ${\cal W}$'s with conformal weights $(1,0)$ ($(0,1)$)
multplied by $\7\xi^+$ ($\7\xi^-$) in order to compensate
for their conformal weight, and $\{y_{+-}^\Delta\}$
contains all ${\cal W}$'s with conformal weights $(1,1)$
multiplied by $\7\xi^+\7\xi^-$.
Note that the subscripts $+$ or $-$ of the $y$'s indicate that
they contain $\7\xi^+$ or $\7\xi^-$.
As the $\7\xi^\mu$
anticommute, the product of two $y$'s vanishes
whenever they have a subsript $+$ or $-$ in common.
This allows us to write the most general 
local $\omega({\cal W})$ with vanishing conformal weights in the form
\begin{equation}
\omega={\cal F}(z)+y_+^A\, {\cal G}^+_A(z)
+y_-^A\, {\cal G}^-_A(z)
+y_{+-}^\Delta{\cal H}_\Delta(z)+y_+^Ay_-^B {\cal K}_{AB}(z)
\label{om}
\end{equation}
where ${\cal F}(z)$, ${\cal G}^\pm_A(z)$, ${\cal H}_\Delta(z)$, 
${\cal K}_{AB}(z)$ are
polynomials in the Grassmann odd
quantities $\7\Theta_+$, $\7\Theta_-$ and $\7C^a$ 
with $X$-dependent coefficient
functions. The remainder of the cohomological analysis
is simply a direct analysis of the cocycle condition 
$\7s\, \omega=0$ with $\omega$ as in (\ref{om}), modulo
corresponding coboundary terms. To perform it we need to determine
first the classical action in order to complete the 
$\7s$-transformations
of $\7X^*_{M+-}$, $\7A^{\mp *}_{a \pm}$ 
and $\7A^{\mu *}_{a +-}$.

\mysection{Action}\label{action}

The classical action has vanishing ghost number and does not depend
on antifields. 
The requirement that the action be gauge invariant is in
our case equivalent to its BRST invariance
(up to surface terms, of course) as
the gauge algebra is closed. The most general gauge invariant local action 
for the models under study is therefore obtained from the
$\7s$-cohomology at total degree 2 in the space of antifield
independent local total forms. We will now compute this cohomology
group and then extract from the result the most general local action 
with the required properties. 

Using the results of the previous section this
is actually an easy exercise. We will nevertheless describe
it in some detail as the procedure applied here will be also 
used later in the more compact notation (\ref{compact})
to complete the cohomological analysis.
{}From the analysis of the previous section we know that
we can restrict ourselves to antifield independent
local total forms (\ref{om}) in order to classify the gauge invariant local
action functionals.
Such total forms thus depend only
on the quantities listed in table 2. The $\7s$-transformations
given in the table
are obtained from the formulae of the previous section.
\[
\begin{array}{c||c||c|c|c||c|c|c}
Q & X^M & \7X^M_\pm & \7\Theta_\pm & \7C^a & \7X^M_{+-} & 
\7Y_\pm & \7F^a_{+-}
\rule{0em}{3ex}\\
\hline
\mbox{totdeg}(Q)  & 0   & 1 & 1 & 1 & 2 & 2 & 2 \\
\hline\rule{0em}{3ex}
\7s\, Q &\7X^M_++\7X^M_-& \mp \7X^M_{+-}& \7Y_\pm &\7F^a_{+-} &0&0&0
\\
\multicolumn{8}{c}{}
\\
\multicolumn{8}{c}{\mbox{Table 2: 
Properties of the antifield independent $z$'s and $y$'s}}
\end{array}
\]
The most general antifield independent
total form (\ref{om}) with total degree 2 is
\begin{eqnarray*}
\omega^2&=& \7X^M_+\7X^N_-f_{MN}(X)+
\7\Theta_+\7\Theta_-f(X)+\7C^a \7C^b f_{ab}(X)
\nonumber\\
     & & +\7X^M_\mu\7\Theta_\nu f_M^{\mu\nu}(X)
         +\7X^M_\mu\7C^a f_{Ma}^\mu(X)
         +\7\Theta_\mu\7C^a f^\mu_a (X)
\nonumber\\
     & &+\7Y_\mu f^\mu(X)+\7F^a_{+-}f_a (X)+\7X^M_{+-} f_M(X)
\end{eqnarray*}
where the $f$'s are arbitrary functions of the $X$'s.
Using table 2, 
one easily computes $\7s\,\omega^2$ and derives
that the cocycle condition 
$\7s\,\omega^2=0$ imposes the following restrictions
on the target space functions occurring in $\omega^2$:
\begin{eqnarray} 
&
\7s\,\omega^2=0 \quad \Leftrightarrow \quad
f=f_{ab}=f^\mu_a=0,\quad 
f_M^{\nu\mu}=\6_M f^\mu,\quad
f_{Ma}^\mu=\6_M h_a
&
\\
& 
\Leftrightarrow \
\omega^2=\7X^M_+\7X^N_-f_{MN}(X)
+(\7X^M_++\7X^M_-)[\7\Theta_\mu\6_M f^\mu(X)+\7C^a\6_M h_a(X)]
&
\nonumber\\
&
+\7Y_\mu f^\mu(X)+\7X^M_{+-} f_M(X)+\7F^a_{+-}f_a(X)
&
\label{act1}\end{eqnarray}
for some functions $h_a(X)$. Here and throughout
the paper we use the notation
\[ \6_M\equiv\frac \6{\6X^M}\ .\]
In order to remove those pieces from $\omega^2$ which are
$\7s$-exact in the space of antifield independent local total
forms, we consider next the general antifield independent
total form (\ref{om}) with total degree 1,
\[
\omega^1= \7X^M_\mu d^\mu_M(X)+\7\Theta_\mu d^\mu(X)+\7C^a d_a (X).
\]
Using table 2 again, one gets
\begin{eqnarray}
\7s\,\omega^1&=&
\7X^M_+\7X^N_-[\6_Md^-_N(X)-\6_Nd^+_M(X)]
\nonumber\\
& &
+(\7X^M_++\7X^M_-)[\7\Theta_\mu\6_M d^\mu(X)
+\7C^a\6_M d_a(X)]
\nonumber\\
& &
+\7Y_\mu d^\mu(X)
+\7X^M_{+-} [d^-_M(X)-d^+_M(X)]+\7F^a_{+-}d_a(X).
\label{act2}
\end{eqnarray}
{}From (\ref{act1}) and (\ref{act2}) we see that by
subtracting the $\7s$-exact piece $\7s\,\omega^1$ 
from an $\7s$-cocycle $\omega^2$, we can
shift some of the target space functions 
occurring in this cocycle,
\begin{equation}
\omega^2 \rightarrow \omega^2-\7s\,\omega^1 \ 
\Leftrightarrow \left\{ \begin{array}{ccl}
f^\mu    & \rightarrow & f^\mu-d^\mu \\
f_M      & \rightarrow & f_M-(d^-_M-d^+_M) \\
\6_M h_a& \rightarrow & \6_M (h_a-d_a) \\
f_a     & \rightarrow & f_a-d_a \\
f_{MN}   & \rightarrow & f_{MN}-(\6_M d^-_N-\6_N d^+_M).
\end{array}\right. 
\label{act3}
\end{equation}
This shows that we can always remove $ f^\mu $, $f_M$
and either $h_a$ or $f_a$ from $\omega^2$ (but, in general,
not both $h_a$ and $f_a$) by
subtracting an $\7s$-exact local total form from $\omega^2$.
We choose
\begin{equation} d^\mu(X)=f^\mu(X), \ 
d^-_M(X)-d^+_M(X)=f_M(X),\ \6_M d_a(X)=\6_M h_a(X).
\label{act5}\end{equation}
Note that this fixes $d^\mu$, $d^\mu_M$ and $d_a$ 
up to arbitrary
functions $e_M=\frac 12 (d^-_M+d^+_M)$ and
constant contributions $e_a$ to the $d_a$.
The choice (\ref{act5}) shifts also the
remaining functions $f_{MN}$ and $f_a$ 
according to (\ref{act3}). We denote these
remaining (shifted) functions by $G_{MN}$, $B_{MN}$ and
$D_a$ respectively, where $G_{MN}$ and $B_{MN}$ are
the symmetric and antisymmetric part of the shifted
$f_{MN}$ respectively.
We conclude that, up to trivial ($\7s$-exact) contributions, 
any local and antifield independent
$\7s$-cocycle with total degree 2 can be chosen 
to be of the form
\begin{equation}
\omega^2=\7X^M_+\7X^N_-[G_{MN}(X)+B_{MN}(X)]
+\7F^a_{+-}D_a(X)
\label{act7}
\end{equation}
where $G_{MN}$ and $B_{MN}$ are symmetric and antisymmetric
in their indices respectively, and we still have the freedom
to shift $B_{MN}$ and $D_a$ according to
\begin{eqnarray}
B_{MN}(X) &\rightarrow & B_{MN}(X)-\6_{[M}e_{N]}(X),
\nonumber\\
D_a(X) &\rightarrow &  D_a(X)-e_a
\label{act8}
\end{eqnarray}
without changing the cohomological class of $\omega^2$.

The 2-form $\omega_2=d\sigma^+d\sigma^- L_0$ 
contained in $\omega^2$ is now easily extracted
from it,
\begin{equation}
L_0=(1-h_{++}h_{--})[G_{MN}(X)+B_{MN}(X)]{\cal D}_+X^M\cdot{\cal D}_-X^N
+\frac 12\, \epsilon^{\mu\nu}D_a(X)F_{\mu\nu}^a.
\label{LBel}
\end{equation}
We conclude that the integrand of any action
with the required properties is, up to a total
derivative, of the form (\ref{LBel}). In terms of the
original fields it reads
\begin{eqnarray}
L_0&=&\frac 12\, \sqrt \gamma\, \gamma^{\mu\nu}
G_{MN}(X)\6_\mu X^M\cdot\6_\nu X^N
\nonumber\\
& &+\frac 12\,\epsilon^{\mu\nu}[B_{MN}(X)\6_\mu X^M\cdot\6_\nu X^N
+D_a(X)F^a_{\mu\nu}].
\label{L}\end{eqnarray}
It is easy to verify that the redefinitions (\ref{act8})
indeed change $L_0$ only by a total derivative.

Recall that we have derived the above result only by fixing the
field content and by requiring locality and gauge invariance 
under world-sheet
diffeomorphisms, local Weyl transformations of the world-sheet metric,
and abelian gauge transformations of the $A^a_\mu$. It is
of course obvious that any action with a Lagrangian (\ref{L})
has these properties. The nontrivial result we have derived
here is that this form of the Lagrangian is unique up to
total derivatives. 

The cohomological analysis carried out in the remainder of this paper
will be performed without specifying the 
functions $G_{MN}(X)$, $B_{MN}(X)$ or $D_a(X)$, i.e. it
applies to almost any model with a Lagrangian of the form (\ref{L}).
We only have to exclude special choices of $G_{MN}(X)$, 
$B_{MN}(X)$ and $D_a(X)$
yielding models with even more gauge symmetries than we have imposed.
For such models our approach would not be sufficient because one
would have to add extra ghost fields and their antifields 
corresponding to the additional gauge symmetries. We shall
discuss models which are excluded for this reason 
in appendix \ref{appA}.

Apart from requiring the absence of extra gauge symmetries
our analysis of models with a Lagrangian of the form (\ref{L})
will be truely general, i.e.\ we will not impose further
restrictions on $G_{MN}$, $B_{MN}$ or $D_a$.
In particular we do not assume that $G_{MN}(X)$ is invertible,
i.e.\ we do not assume that it can be viewed as a metric of a target
space of all $X^M$. Among others this allows us to include
bosonic (effective) D-string actions of the Born--Infeld type.
Indeed, such actions arise from (\ref{L}) for
specific choices of $G_{MN}$, $B_{MN}$ or $D_a$
with degenerate $G_{MN}$ \cite{letter}. To give an example,
we single out one target space coordinate
$\varphi$ and use the notation
\begin{equation} 
X^0=\varphi,\quad X^M=x^m\quad  \mbox{for}\quad m=M>0, 
\label{BI1}
\end{equation}
and choose
\begin{eqnarray*}
& G_{M0}(X)=0,\quad 
G_{mn}(X)=(1-\varphi^2)^{-1/2}e^{-\phi(x)}g_{mn}(x)&
\\
& B_{m0}=0,\quad 
B_{mn}(X)=-\varphi (1-\varphi^2)^{-1/2}e^{-\phi(x)}b_{mn}(x)
          +c_{mn}(x)&
\\
& D_a(X)=-\varphi (1-\varphi^2)^{-1/2}e^{-\phi(x)}d_a(x)+c_a(x). &
\end{eqnarray*}
Eliminating the
auxiliary fields $\varphi$ and $\gamma_{\mu\nu}$,
the Lagrangian (\ref{L}) turns indeed 
into the Born--Infeld type Lagrangian
\begin{eqnarray}
& 
L_{BI}=e^{-\phi(x)}\sqrt{-\det ({\cal G}_{\mu\nu}+{\cal F}_{\mu\nu})}
+\frac 12\, \epsilon^{\mu\nu}
\left[ c_a(x)F_{\mu\nu}^a+c_{mn}(x)\6_\mu x^m \6_\nu x^n \right]
&
\nonumber\\
& {\cal G}_{\mu\nu}=g_{mn}(x)\6_\mu x^m \6_\nu x^n &
\nonumber\\
& {\cal F}_{\mu\nu}=d_a(x)F_{\mu\nu}^a+b_{mn}(x)\6_\mu x^m \6_\nu x^n.&
\label{BI4}
\end{eqnarray}
Of course, the `dilaton factor' $\exp(-\phi(x))$ can be absorbed 
into $g_{mn}(x)$, $b_{mn}(x)$ and $d_a(x)$, analogously to the 
switch from the `string frame' to the `Einstein frame'.
	
There are further potentially interesting actions
of this type associated with
different choices of $G_{MN}$, $B_{MN}$ and $D_a$.  
For example, let us consider (\ref{BI1}) and
\begin{eqnarray*}
& G_{M0}(X)=0,\quad G_{mn}(X)=(1+\varphi^2)g_{mn}(x)&
\\
& B_{m0}=0,\quad 
B_{mn}(X)=2\varphi\, b_{mn}(x),\quad
D_{a}(X) = 2\varphi\, d_{a}(x).&
\end{eqnarray*}
If we eliminate the auxiliary fields $\varphi$ and
$\gamma_{\mu\nu}$ we get the Lagrangian
\begin{equation}
L =\frac{-\det({\cal G}_{\mu\nu}+{\cal F}_{\mu\nu})}
{\sqrt{-\det({\cal G}_{\mu\nu})}}\ .
\end{equation}

\mysection{Result}\label{Result}

{}From (\ref{L}) one obtains the Koszul--Tate part of the 
BRST transformations
of the antifields $\phi^*_i$ in (\ref{delta}),
\begin{eqnarray}
\delta X^*_M &=& 
-G_{MN} \6_\mu (\sqrt \gamma \gamma^{\mu\nu}\6_\nu X^N)
-\sqrt \gamma \gamma^{\mu\nu}\Gamma_{KLM} \6_\mu X^K\cdot\6_\nu X^L
\nonumber\\
& &+\frac 12\, \epsilon^{\mu\nu}( F^a_{\mu\nu} \6_MD_a
+H_{KLM} \6_\mu X^K\cdot\6_\nu X^L)
\nonumber\\
\delta A^{\mu *}_a &=&\epsilon^{\mu\nu}\6_\nu D_a 
\nonumber\\
\delta \gamma^{\mu\nu *} &=&
\sqrt \gamma\, G_{MN}(\frac 14 \gamma^{\mu\nu}
\gamma^{\rho\sigma}\6_\rho X^M\cdot\6_\sigma X^N
-\frac 12\6^\mu X^M\cdot\6^\nu X^N)
\label{brs6}\end{eqnarray}
with $\6^\mu\equiv \gamma^{\mu\nu}\6_\nu$, and 
\begin{eqnarray}
\Gamma_{KLM}&=&\frac 12\, (\6_KG_{LM}+\6_LG_{KM}-\6_MG_{KL}),
\nonumber\\
H_{KLM}&=&\6_KB_{LM}+\6_LB_{MK}+\6_MB_{KL}\ .
\label{GH}\end{eqnarray}
This yields the $\7s$-transformations in table 3.
\[
\begin{array}{c|c|c}
Q           & \mbox{totdeg} (Q) & \7s\, Q 
\rule{0em}{3ex}\\
\hline
X^M         &    0     & \7X^M_++\7X^M_-
\rule{0em}{3ex}\\
\7\Theta_\pm  &    1     & \7Y_\pm 
\rule{0em}{3ex}\\
\7C^a      &    1     & \7F^a_{+-} 
\rule{0em}{3ex}\\
\hline
\7A_{a\pm}^{\mp *}
	    &    0     & \pm (\7X^M_\pm\6_M D_a-\7A_{a +-}^{\mp *})
\rule{0em}{3ex}\\
\7X^M_\pm     &    1     & \mp \7X^M_{+-}
\rule{0em}{3ex}\\
\7Y_\pm       &    2     & 0
\rule{0em}{3ex}\\
\hline
\7C^*_{a +-} &    0     & -\7A_{a +-}^{+ *}-\7A_{a +-}^{- *}
\rule{0em}{3ex}\\
\7A_{a +-}^{\pm *}
	    &    1     & \pm (\7X^M_+\7X^N_-\6_M\6_N D_a+\7X^M_{+-}\6_M D_a)
\rule{0em}{3ex}\\
\7X_{M+-}^*   &    1     & -2G_{MN}\7X^N_{+-}
			 +(H_{KLM}-2\Gamma_{KLM}) \7X^K_+ \7X^L_-
\rule{0em}{3ex}\\
              &          & +\7F^a_{+-}\6_M D_a
\rule{0em}{3ex}\\
\7X^M_{+-}    &    2     & 0
\rule{0em}{3ex}\\
\7F_{+-}^a   &    2     & 0
\\
\multicolumn{3}{c}{\mbox{Table 3: 
Total degrees and $\7s$-transformations
of the $z$'s and $y$'s}}
\rule{0em}{5ex}
\end{array}
\]

The cohomology can now be computed directly by applying $\7s$ to the 
most general local total form (\ref{om}), and elaborating
the conditions imposed by $\7s$ invariance and nontriviality.
These are conditions on the functions of the $z$'s occurring 
in (\ref{om}), i.e.\ the aim is to determine these functions.
The computation is analogous to the one performed in the
previous section to determine the action functionals, but of course
more involved. We have relegated it to appendix \ref{appB}
and will summarize the results in the following.

In order to present these results we
use a ``superfield notation''.
To this end we expand the functions of the $z$'s in terms of
$\7\Theta_+$ and $\7\Theta_-$. As the $\7\Theta$'s anticommute,
this expansion contains only a few terms and is indeed reminiscent
of a superfield. Such ``superfields'' are always denoted by
calligraphic  capital letters while we use corresponding
small latin letters for their ``component functions'', such as
\begin{equation} 
{\cal F}(z)=f(X,\7C)+\7\Theta_\mu f^\mu(X,\7C)
+\7\Theta_+\7\Theta_-\5 f(X,\7C).
\label{superfield}\end{equation}
Note that
the component functions depend on the $X^M$ and $\7C^a$, 
i.e.\ they involve both bosonic and fermionic variables.
For notational convenience, we shall often
leave out the arguments of the superfields and their
component functions.

Using this notation, the result can be described as follows.
Modulo $\7s$-exact contributions, the most general 
$\7s$-invariant local total form is
\begin{eqnarray}
\omega   &=& \7F^{a}_{+-}{\cal H}_{a}
   + \7X_{+}^{M}\7X_{-}^{N}{\cal K}_{MN}
   + \7X_{+}^{M}\7Y_{-}{\cal K}_{M-} + \7Y_{+}\7X_{-}^{M}{\cal K}_{+M}
\nonumber\\
& &
   + \7Y_{+}\7Y_{-}{\cal K}
   + \7Y_{+}({\cal G}^{+}+\7A^{+*}_{a -}{\cal K}^{a}_{+}) 
   + \7Y_{-}({\cal G}^{-}+\7A^{-*}_{a +}{\cal K}^{a}_{-})
\nonumber\\
& &
   + \7X^{*}_{M+-}{\cal H}^{M} 
   + (\7C^{*}_{a +-} + \7A^{+*}_{a -} - \7A^{-*}_{a +}){\cal H}^{a} 
   + \7A^{-*}_{a +}\7A^{+*}_{b -}{\cal K}^{ab}
\nonumber \\
& &
    + \frac{1}{2}(\7X_{+}^{M}\7A^{+*}_{a -}
                 +\7X_{-}^{M}\7A^{-*}_{a +}){\cal A}^{a}_{M}
    + \frac{1}{2}(\7X_{+}^{M}\7A^{+*}_{a -}
                 -\7X_{-}^{M}\7A^{-*}_{a +}){\cal B}^{a}_{M}
\nonumber \\
& &
    + (\7X_{-}^{M} - \7X_{+}^{M})G_{MN}{\cal H}^N
    + \frac{1}{2}(\7X_{+}^{M} + \7X_{-}^{M}){\cal P}_{M} 
    + {\cal F}\ .
\label{gs1}
\end{eqnarray}
Here ${\cal H}_{a}$ and ${\cal K}_{MN}$ are arbitrary
``superfields'',
while the other ones have the following expansions in component functions,
\begin{eqnarray}
{\cal K}_{M-} &=& \7\Theta_+k_M+\7\Theta_-k_{M-}^-
            +\7\Theta_+\7\Theta_-\5k_{M-}
\nonumber\\
{\cal K}_{+M} &=& \7\Theta_-k_M+\7\Theta_+k_{+ M}^+
            +\7\Theta_+\7\Theta_-\5k_{+ M}
\nonumber\\ 
{\cal K}\ &=& \7\Theta_+\7\Theta_-\5k
\nonumber\\
{\cal K}^{a}_\pm &=& \7\Theta_\pm k_{\pm}^{a\pm}
\nonumber\\
{\cal G}^{\pm} &=& \7\Theta_\pm\, g^{\pm\pm}
\nonumber\\
{\cal H}^M &=& h^M+\7\Theta_+h^{M+}+\7\Theta_-h^{M-}
\nonumber\\
{\cal A}^a_M &=& a^a_M+\7\Theta_+a^{a+}_M+\7\Theta_-a^{a-}_M
\nonumber\\
{\cal B}^a_M &=& b^a_M+\7\Theta_+b^{a+}_M+\7\Theta_-b^{a-}_M
\nonumber\\
{\cal P}_M &=& p_M+2G_{MN}(\7\Theta_- h^{N-}-\7\Theta_+ h^{N+})
\nonumber\\
{\cal H}^{a} &=& h^a,\quad {\cal K}^{ab}= k^{(ab)},\quad {\cal F} = f,
\label{result}\end{eqnarray}
and have to satisfy
\begin{eqnarray}
& & \partial_{(M}[G_{N)K}{\cal H}^K] - \Gamma_{MNK}{\cal H}^{K} + 
\frac{1}{2}{\cal A}^{a}_{(M}\partial_{N)}D_{a} = 0
\label{result1}\\
& & \partial_{[M}{\cal P}_{N]} + H_{MNK}{\cal H}^{K} + 
{\cal B}^{a}_{[M}\partial_{N]}D_{a} = 0
\label{result2}\\
& & \frac{\partial{\cal F}}{\partial\7C^{a}} 
   + {\cal H}^{M}\partial_{M}D_{a}=0
\label{result3a}\\
& &   \partial_{M}{\cal F} - {\cal H}^{a}\partial_{M}D_{a}=0
\label{result3b}\\
& & \partial_{M}{\cal H}^{a} + {\cal K}^{ab}\partial_{M}D_{b} = 0
\label{result4}\\
& & \partial_{M}{\cal G}^{\mp} \pm {\cal K}^{a}_{\mp}\partial_{M}D_{a}=0.
\label{result5}
\end{eqnarray}

One can still subtract trivial ($\7s$-exact) local total forms
$\7s\,\hat\omega$ from $\omega$ without changing the form of
$\omega$. These total forms $\hat\omega$ are of the form
\begin{eqnarray}
\hat\omega &=& 
\hat f+\frac 12(\7X_{+}^{M} + \7X_{-}^{M})\hat{\cal P}_{M}
+(\7X_{-}^{M} - \7X_{+}^{M})[G_{MN}\hat {\cal H}^N
-\hat {\cal Z}^a \6_M D_a ]
\nonumber\\
& &
   +\7Y_+(\hat g^+  + \7\Theta_+\hat g^{++})
   +\7Y_-(\hat g^-  + \7\Theta_-\hat g^{--})
\nonumber\\
& &
   + \7X^{*}_{M+-}\hat {\cal H}^{M} 
   + (\7C^{*}_{a +-} + \7A^{+*}_{a -} - \7A^{-*}_{a +})\hat h^{a} 
\nonumber \\
& &
   +\7X_{+}^{M}\7A^{+*}_{a -}\hat {\cal K}_M{}^a
   + \7X_{-}^{M}\7A^{-*}_{a +}\hat {\cal K}^a{}_M
   + \7Y_+ \7A^{+*}_{a -}\hat {\cal K}_+^a 
   + \7A^{-*}_{a +}\7Y_-\hat {\cal K}_-^a
\nonumber \\
& &
   + (\7A^{+*}_{a +-} - \7A^{-*}_{a +-})\hat {\cal Z}^a
   + \7A^{-*}_{a +}\7A^{+*}_{b -}\hat {\cal K}^{ab}
\label{hatomega}\end{eqnarray}
where $\hat {\cal P}_M$ and $\hat {\cal K}^{ab}$ are given by
\begin{eqnarray}
\hat {\cal P}_M &=& \hat p_M+\7\Theta_\mu\hat p_M^\mu
              +\7\Theta_+\7\Theta_-\overline{\hat p}_M 
\label{choice14a}\\
\hat p_M^\pm &=& \mp 2G_{MN}\hat h^{N\pm}+2\6_M\hat g^\pm
\pm 2(\hat z^{a\pm}-\hat k_{\pm}^a)\6_M D_a
\label{choice14b}\\
\overline{\hat p}_M &=& 
-(\hat k^{a+}_{-}+\hat k_{+}^{a-})\6_M D_a 
\label{choice14c}
\\
\hat {\cal K}^{ab} &=& \hat k^{ab}+\7\Theta_\mu\hat k^{ab\mu}.
\label{choice14d}\end{eqnarray}
In fact, up to $\7s$-invariant contributions, $\hat\omega$
is the most general local total form 
such that subtracting $\7s\,\hat\omega$
from $\omega$ only modifies the component 
functions of the superfields appearing in (\ref{gs1}) and (\ref{result}),
without introducing additional ones.
A shift $\omega  \rightarrow  \omega - \7s\,\hat\omega$
results in
\begin{eqnarray}
{\cal P}_M & \rightarrow & {\cal P}_M -2\6_M\hat f+2\hat h^a\6_M D_a
\label{shifts1}\\
{\cal A}_M^a & \rightarrow & {\cal A}_M^a +2\hat {\cal K}^{[ab]}\6_M D_a 
\label{shifts2}\\
{\cal B}_M^a & \rightarrow & {\cal B}_M^a -2\6_M\hat h^a
                        - 2\hat {\cal K}^{(ab)}\6_M D_a 
\label{shifts3}\\
{\cal H}_a & \rightarrow & {\cal H}_a - \frac{\6\hat f}{\6 \7C^a}
                        - \hat {\cal H}^M\6_M D_a 
\label{shifts4}\\
{\cal K}_{MN} & \rightarrow & {\cal K}_{MN} 
     - ( \partial_{[M}\hat {\cal P}_{N]} + H_{MNK}\hat {\cal H}^{K} + 
       \hat {\cal B}^{a}_{[M}\partial_{N]}D_{a} ) 
\nonumber\\
 & & - 2( \partial_{(M}[G_{N)K}\hat {\cal H}^K] 
      - \Gamma_{MNK}\hat {\cal H}^{K}
      + \frac {1}{2}\hat {\cal A}^{a}_{(M}\partial_{N)}D_{a} )
\label{shifts6}\\
k_M  & \rightarrow & k_M + G_{MN}\overline{\hat h}{}^N
      - \frac 12(2\overline{\hat z}{}^a
      +\hat k{}_-^{a+}-\hat k{}_+^{a-})\6_M D_a 
\label{shifts7}\\
k_{M-}^- & \rightarrow & k_{M-}^-
                         - \hat k^{a-}_-\6_M D_a - \6_M\hat g^{--} 
\label{shifts8}\\
k_{+M}^+ & \rightarrow & k_{+M}^+
                         - \hat k^{a+}_+\6_M D_a - \6_M\hat g^{++} 
\label{shifts9}\\
\5 k_{+M} & \rightarrow &
            \5 k_{+M}+\overline{\hat k}{}^a_+ \6_M D_a
\label{shifts10a}\\
\5 k_{M-} & \rightarrow &
            \5 k_{M-}-\overline{\hat k}{}^a_- \6_M D_a
\label{shifts10b}
\end{eqnarray}
where 
\begin{eqnarray}
\hat {\cal A}^{a}_{M} &=& 
\hat {\cal K}^{a}{}_M + \hat {\cal K}_M{}^{a}
- 2\partial_{M}\hat {\cal Z}{}^{a}
\nonumber\\
\hat {\cal B}^{a}_{M} &=& \hat {\cal K}_{M}{}^{a} - \hat {\cal K}^{a}{}_{M}\ .
\label{hatAB}
\end{eqnarray}
All other superfields in $\omega$ remain unchanged, i.e.\ they cannot
be removed or modified by subtracting $\7s$-exact contributions
from $\omega$ without changing its form.

(\ref{gs1}) contains many separately
$\7s$-invariant solutions. We group them in eight
types, three of which appear in two ``chiralities''.
Among others, this will facilitate the discussion of the results
later on.
\begin{eqnarray}
\omega   &=& \sum_{i=1,2,3,5,6} \omega_i
          +\sum_{i=4,7,8}(\omega_{i+}+\omega_{i-}),\quad
\7s\, \omega_i=\7s\, \omega_{i\pm}=0;
\label{solutions}
\\
\omega_1 &=& \7F^{a}_{+-}{\cal H}_{a}(z) 
\label{om1}
\\
\omega_2 &=& \7X_{+}^{M}\7X_{-}^{N}{\cal K}_{MN}(z) 
\label{om2}
\\
\omega_3 &=& (\7X_{+}^{M}\7Y_{-}\7\Theta_+ 
+ \7Y_{+}\7X_{-}^{M}\7\Theta_-) k_M(X,\7C)
\label{om3}
\\
\omega_{4+} &=& \7X_{+}^{M}\7Y_{-}
               \left[\7\Theta_-k_{M-}^-(X,\7C)
               +\7\Theta_+\7\Theta_-\5k_{M-}(X,\7C)\right]
\label{om3+}
\\
\omega_{4-} &=& \7Y_{+}\7X_{-}^{M} 
               \left[\7\Theta_+k_{+M}^+(X,\7C)
               +\7\Theta_+\7\Theta_-\5k_{+M}(X,\7C)\right]
\label{om3-}
\\
\omega_5 &=& \7Y_{+}\7Y_{-}\7\Theta_+\7\Theta_-\bar k(X,\7C)
\label{om4}
\\
\omega_6 &=& \7X^{*}_{M+-}h^{M}(X,\7C)
    + (\7C^{*}_{a +-} + \7A^{+*}_{a -} - \7A^{-*}_{a +})h^{a}(X,\7C) 
\nonumber \\
& & 
    + \frac{1}{2}(\7X_{+}^{M}\7A^{+*}_{a -}
                 +\7X_{-}^{M}\7A^{-*}_{a +})a^{a}_{M}(X,\7C)
    + \frac{1}{2}(\7X_{+}^{M}\7A^{+*}_{a -}
                 -\7X_{-}^{M}\7A^{-*}_{a +})b^{a}_{M}(X,\7C)
\nonumber\\
& &
    + \7A^{-*}_{a +}\7A^{+*}_{b -}k^{(ab)}(X,\7C)
    + (\7X_{-}^{M} - \7X_{+}^{M})G_{MN}(X)h^N(X,\7C) 
\nonumber \\
& &
    + \frac{1}{2}(\7X_{+}^{M} + \7X_{-}^{M})p_{M}(X,\7C)+ f(X,\7C) 
\label{om6}
\\
\omega_{7\pm} &=& -\7\Theta_\pm \left[ \7X^{*}_{M+-} h^{M\pm}(X,\7C)
    +\frac{1}{2}(\7X_{+}^{M}\7A^{+*}_{a -}
      +\7X_{-}^{M}\7A^{-*}_{a +}) a^{a\pm}_{M}(X,\7C)
\right.
\nonumber\\
& &\phantom{-\7\Theta_\pm \left[\right.}
    + \frac{1}{2}(\7X_{+}^{M}\7A^{+*}_{a -}
      -\7X_{-}^{M}\7A^{-*}_{a +}) b^{a\pm}_{M}(X,\7C)
\nonumber\\
& &\phantom{-\7\Theta_\pm \left[\right.}\left.
    \mp 2\7X^M_\pm G_{MN}(X)h^{N\pm}(X,\7C)
\right]
\label{om7}
\\
\omega_{8\pm} &=& \7Y_\pm\7\Theta_\pm\left[ g^{\pm\pm}(X,\7C)
   + \7A^{\pm*}_{a \mp}k^{a\pm}_{\pm}(X,\7C)\right]
\label{om5}
\end{eqnarray}
Note that
$\omega_1$, \dots , $\omega_{5}$ do not depend
on antifields. Each of them is $\7s$-invariant for any choice of the
component functions occurring in them. 
In contrast, $\omega_6$,
$\omega_{7\pm}$ and $\omega_{8\pm}$ involve in general antifields, and the
component functions occurring in them are determined by partial
differential equations following from 
(\ref{result1})--(\ref{result5}). Let us now spell out these
equations explicitly for the respective solutions.

(\ref{result1})--(\ref{result4}) impose the following equations
on $\omega_6$,
\begin{eqnarray}
& & \partial_{(M}[G_{N)K}h^K] - \Gamma_{MNK}h^{K} 
+\frac 12\, a^{a}_{(M}\partial_{N)}D_{a} = 0 
\nonumber\\
& & \partial_{[M}p_{N]} + H_{MNK}h^{K} 
+b^{a}_{[M}\partial_{N]}D_{a} = 0
\nonumber\\
& &  \frac{\partial f}{\partial\7C^{a}} 
   + h^{M}\partial_{M}D_{a}=0
\nonumber\\
& &  \partial_{M}f - h^{a}\partial_{M}D_{a} =0
\nonumber\\
& & \partial_{M}h^{a} + k^{(ab)}\partial_{M}D_{b} = 0 
\label{cond6}
\end{eqnarray}
while (\ref{result5}) does not constrain the functions in $\omega_6$.
Eqs. (\ref{cond6}) imply $\7s\, \omega_6=0$.

(\ref{result1})--(\ref{result3a}) require the functions in
$\omega_{7\pm}$ to solve
\begin{eqnarray}
& \partial_{M}(G_{NK}h^{K\pm}) 
- (\Gamma_{MNK}\pm \frac 12 H_{MNK})h^{K\pm} =
-(v^{a\pm}_M\6_N+w^{a\pm}_N\6_M) D_a &
\nonumber\\
&  h^{M\pm}\6_M D_a=0 &
\nonumber\\
&
\mbox{where}\quad  
v^{a\pm}_M = \frac 14 (a^{a\pm}_M \mp b^{a\pm}_M),\quad
w^{a\pm}_M = \frac 14 (a^{a\pm}_M \pm b^{a\pm}_M).
&
\label{cond7}
\end{eqnarray}
In (\ref{cond7}), we have merged the two equations 
originating from (\ref{result1}) and (\ref{result2})
to a single one. (\ref{result3b})--(\ref{result5}) 
do not constrain $\omega_{7\pm}$.
Eqs. (\ref{cond7}) imply $\7s\, \omega_{7\pm}=0$.

(\ref{result1})--(\ref{result4}) do not constrain $\omega_{8\pm}$.
The only conditions imposed by
(\ref{result5}) on $\omega_{8\pm}$ are
\begin{equation}
\6_M g^{\pm\pm} = \pm k_{\pm}^{a\pm}\6_M D_a\ .
\label{cond5}\end{equation}
This implies $\7s\, \omega_{8\pm}=0$.

Let us finally discuss whether $\omega_6$,
$\omega_{7\pm}$ or $\omega_{8\pm}$ provide also antifield independent
solutions. Clearly, $\omega_{8\pm}$ does not depend on antifields only
if $k^{a\pm}_{\pm}=0$. (\ref{cond5}) implies in this case 
$\6_M g^{\pm\pm}=0$, and thus $g^{\pm\pm}=g_0^{\pm\pm}(\7C)$. 
These solutions are nontrivial because $g^{\pm\pm}$ cannot
be redefined by subtracting trivial solutions from $\omega$ (it
does not appear in (\ref{shifts1})--(\ref{shifts10b})).
Hence, the antifield independent solutions arising
from $\omega_{8\pm}$ are
\begin{equation} \stackrel{o}{\omega}_{8\pm}=\7Y_\pm\7\Theta_\pm
g_0^{\pm\pm}(\7C).
\label{om5o}
\end{equation}

$\omega_{6}$ does not depend on antifields only if
$h^M=h^a=a^a_M=b^a_M=k^{(ab)}=0$. In this case, (\ref{cond6})
implies $f=constant$ and $p_M=\6_M p$ for some $p(X,\7C)$.
(\ref{shifts1}) shows that we can assume $\6_M p=0$ without
loss of generality (by choosing $\hat f$ appropriately). Hence,
the only nontrivial antifield independent solutions arising 
from $\omega_{6}$ are the constants. They can of course be neglected
as they do not provide solutions to (\ref{i1}).

$\omega_{7+}$ and $\omega_{7-}$ do not provide antifield independent
solutions at all.

\mysection{Discussion of the results}\label{discussion}

We will now spell out and discuss the physically most
important results of our analysis, i.e.\ the inequivalent
solutions of (\ref{i1}) with ghost numbers $g\leq 1$.
They provide the rigid symmetries, dynamical
conservation laws, gauge invariant actions, the
first order consistent deformations of these actions 
and of their gauge symmetries, and the candidate gauge anomalies.

As explained in section \ref{contractions},
the integrands of the nontrivial 
BRST invariant local functionals with ghost number $g$ are
the 2-forms contained in the nontrivial total forms 
(\ref{om1})--(\ref{om5}) with total degree $g+2$. 
In order to extract the results 
at ghost number $g$, we use in 
(\ref{om1})--(\ref{om5}) expansions in the $\7C^a$ such as
\begin{equation}
f(X,\7C)=\sum_{m\geq 0}\frac 1{m!}\,\7C^{a_1}\cdots \7C^{a_m}
                            f_{a_1\cdots a_m}(X).
\label{decomposition}
\end{equation}
Note that we use the same letter $f$
for the ghost independent part $f(X)$ as for the 
whole function $f(X,\7C)$.

Recall that the antifield dependent total forms (\ref{om6})--(\ref{om5})
involve component functions which have to satisfy the
partial differential equations (\ref{cond6})--(\ref{cond5})
respectively. Whether or not these equations have nontrivial
solutions depends of course on the particular model, i.e.\ on the
choice of $G_{MN}$, $B_{MN}$ and $D_a$ in the
Lagrangian (\ref{L}). Therefore it also depends on the model whether and
which antifield dependent solutions to (\ref{i1}) exist. In contrast,
the antifield independent solutions do not depend on
$G_{MN}$, $B_{MN}$ and $D_a$, but notice that it nevertheless 
depends on the model
which of these solutions are nontrivial or inequivalent, as 
$G_{MN}$, $B_{MN}$ and $D_a$ enter the 
cohomologically trivial redefinitions
(\ref{shifts1})--(\ref{shifts10b}). In particular, solutions
which are equivalent in one model
might not be equivalent in another one.

\subsection{{\boldmath $g=-2$}: Dynamical conservation laws
of second order}\label{2ndorder}

Nontrivial solutions to (\ref{i1}) with ghost number $-2$ 
arise solely from the total forms (\ref{om6}) with vanishing total degree. 
These total forms involve only the ghost independent parts
of $h^a$, $k^{(ab)}$ and $f$.
The corresponding solutions
to (\ref{i1}) are
\begin{equation}
W^{-2}=\int d^2\sigma\left[C^*_a h^a(X)-
\frac 12\, \epsilon_{\mu\nu}
A^{\mu *}_a A^{\nu *}_b k^{(ab)}(X)\right]
\label{W-2}\end{equation}
where $h^a(X)$ and $k^{(ab)}(X)$ have to satisfy
\begin{equation}
\6_M h^a(X)+k^{(ab)}(X)\6_M D_b(X)=0.
\label{secsymm}
\end{equation}
(\ref{secsymm}) is obtained from (\ref{cond6}) and guarantees
that $W^{-2}$ is BRST invariant. 

Note that $f(X)$ does not occur in (\ref{W-2}), although it
enters the corresponding total form (\ref{om6}). The reason
is that it contributes only to the 0-form in the
descent equations (\ref{descent}) corresponding to $W^{-2}$. 
(\ref{cond6}) relates $f(X)$ to $h^a(X)$ through
\begin{equation}
\6_M f(X)=h^a(X)\6_M D_a(X).
\label{secsymm2}
\end{equation}
As (\ref{secsymm}) guarantees already the BRST invariance
of $W^{-2}$, and as the BRST invariance of $W^{-2}$
implies in turn the descent equations, any solution to
(\ref{secsymm}) must imply the existence of a
corresponding solution to (\ref{secsymm2}). To verify this, we
contract (\ref{secsymm}) with $\6_N D_a$
and antisymmetrize afterwards in $M$ and $N$. This yields
\[
\6_{[M}(h^a\6_{N]} D_a)=0\ \Rightarrow\ 
\exists\, f:\ 
h^a\6_M D_a=\6_M f
\]
where we used the ordinary Poincar\'e lemma for closed 1-forms 
(on the target space), i.e.\ $\6_{[M}V_{N]}=0$ $\Rightarrow$
$V_M=\6_M f$ with $V_M=h^a\6_M D_a$.

Conversely, any solution to (\ref{secsymm2}) implies 
the existence of a corresponding solution to (\ref{secsymm}).
Indeed, differentiating (\ref{secsymm2}) with respect to
$X^N$ and antisymmetrizing afterwards in $M$ and $N$ yields
\begin{equation}
\6_{[M}h^a\6_{N]} D_a=0\  
\Rightarrow\ \exists\, k^{(ab)}:\ \6_M h^a+k^{(ab)}\6_M D_b=0.
\label{then}\end{equation}
Here we have used (\ref{extra1a})--(\ref{extra2}), 
setting there $\eta^M=\lambda^a=a_M^a=0$
and $b_M^a=\6_M h^a$.

Hence, (\ref{secsymm}) and (\ref{secsymm2}) are actually 
equivalent conditions. This is in fact
not surprising as it illustrates the general feature that the
cohomology classes of the
local BRST cohomology at negative ghost numbers $g=-k$ correspond
one-to-one to the dynamical local
conservation laws of order $k$ \cite{bbh1}. In $n$ dimensions,
the latter are represented
by weakly (= on-shell) closed $(n-k)$-forms $j_{n-k}$ 
($d\, j_{n-k}\approx 0$)
which are not weakly exact {\em locally} (in contrast,
{\em topological} conservation laws are locally but not
globally $d$-exact), or, in dual notation, by
completely antisymmetric generalized currents
$j^{\mu_1\cdots\mu_k}$ solving 
\[ \6_{\mu_1}\, j^{\mu_1\cdots\mu_k}\approx 0,\quad
j^{\mu_1\cdots\mu_k}\not\approx 
\6_{\mu_0}\,\hat \jmath^{\,[\mu_0\cdots\mu_k]}.
\]
Now, (\ref{secsymm2}) implies that $f(X)$ is a 
dynamical conservation law of second
order ($k=2$), as it gives 
\begin{equation} d\, f(X)\approx 0 \ \Leftrightarrow\ 
\6_\mu\, j^{\mu\nu}\approx 0,\ 
j^{\mu\nu}=\epsilon^{\mu\nu} f(X).
\label{secsymm3}
\end{equation} 
This is obtained
by contracting (\ref{secsymm2}) with $d X^M$ due to
$d D_a\approx 0$. Note also that $f(X)$ cannot be
trivial as it does not contain derivatives (whereas all
equations of motions do contain derivatives).
We conclude that the nonconstant
solutions $f(X)$ to (\ref{secsymm2}) exhaust
the nontrivial dynamical conservation laws of second order.

It is easy to verify that there are always infinitely many
second order conservation laws. Indeed, as we have just mentioned,
the functions $D_a(X)$ themselves are conserved. Therefore any
function $Z(D(X))$ of the $D_a(X)$ is conserved too. Solutions to
(\ref{secsymm}) and  (\ref{secsymm2}) are thus given by
\begin{equation}
f=Z(D),\quad h^a=\frac{\partial Z}{\partial D_a},\quad
k^{(ab)}=-\frac{\partial^2 Z}{\partial D_a\partial D_b}\ .
\label{infty2nd}
\end{equation}

The solutions $W^{-2}$ themselves have an interpretation too:
as explained in \cite{bhw}, they generate
rigid symmetries $\Delta_{(2)}$ of the {\em extended} action ${\cal S}$ 
(= proper solution of the master equation)
through $\Delta_{(2)}\,\cdot\,=(W^{-2},\,\cdot\,)$.
These symmetries act
nontrivially only on $C^a$, $A_\mu^a$ and $X^*_M$ according to
\begin{eqnarray}
& \Delta_{(2)}C^a = -h^a(X),\quad
\Delta_{(2)}A_\mu^a = -\epsilon_{\mu\nu}A^{\nu *}_b k^{(ab)}(X)&
\nonumber\\
& 
\Delta_{(2)}X^*_M = C^*_a\6_M h^a(X)
-\frac 12\, \epsilon_{\mu\nu}A^{\mu *}_a A^{\nu *}_b\6_M k^{(ab)}(X).
&
\label{Delta2}
\end{eqnarray}

\subsection{{\boldmath $g=-1$}: Rigid symmetries and Noether currents}

The nontrivial solutions to (\ref{i1}) 
with ghost number $-1$ arise also solely from (\ref{om6}) where
this time we have to pick the contribution with total degree 1.
The latter contains those pieces of $h^a$, $k^{(ab)}$ and $f$
which are linear in the $\7C^a$, i.e.\ 
$\7C^b h^a_b(X)$, $\7C^c k^{(ab)}_c(X)$ and
$\7C^a f_a(X)$, as well as
the ghost independent pieces of $h^M$, $a_M^a$,
$b_M^a$ and $p_M$. This
yields the following BRST invariant functionals
\begin{eqnarray}
W^{-1}=\int d^2\sigma\, \left\{
\frac 12\, A^{\mu *}_a [b^a_{M}(X)\6_\mu X^M-\sqrt{\gamma}\, 
\epsilon_{\mu\nu}
a^a_M(X)\6^\nu X^M+2A^b_\mu h^a_b(X)]\right.
\nonumber\\
\left.
+X^*_M h^M(X)+C^*_a C^b h_b^a(X)
+\frac 12\,\epsilon_{\mu\nu}A^{\nu *}_a A^{\mu *}_b C^c k_c^{(ab)}(X)
\right\}
\label{W-1}\end{eqnarray}
where $\6^\mu=\gamma^{\mu\nu}\6_\nu$ and the target space functions have to
solve Eqs.\ (\ref{cond6}) which we will discuss now in some 
more detail.
First we note that one of these equations determines $f_a(X)$ 
in terms of $h^M(X)$ and $D_a(X)$,
\begin{equation}
\frac{\6f(X,\7C)}{\6\7C^a}+h^M(X,\7C)\, \6_M D_a(X)=0\ \Rightarrow\
f_a(X)=-h^M(X)\, \6_M D_a(X).
\label{fa}
\end{equation}
The other Eqs.\ (\ref{cond6}) give now equations for the
remaining target space functions which can be cast in the form
\begin{eqnarray}
 {\cal L}_h G_{MN}(X) &=& -a_{(M}^a(X)\, \6_{N)} D_a(X) 
\label{lie1}\\
 {\cal L}_h B_{MN}(X) &=& -\6_{[M}\, p'{}_{N]}(X)-b_{[M}^a(X)\, \6_{N]} D_a(X)
\label{lie2}\\
 {\cal L}_h \6_M D_a(X) &=& -h_a^b(X)\, \6_M D_b (X)
\label{lie3}\\
 \6_M h^b_a(X) &=& -k^{(bc)}_a(X)\, \6_M D_c (X)
\label{lie4}
\end{eqnarray}
where ${\cal L}_h$ denotes the standard Lie derivative along $h^M(X)$, such as
\[ {\cal L}_h G_{MN}=h^K\6_K G_{MN}
+\6_M h^K\cdot G_{KN}+\6_N h^K\cdot G_{MK},\]
and $p'{}_M(X)$ is defined by
\begin{equation}
p'{}_M(X)=p_M(X)+2B_{MN}(X)\, h^N(X).
\label{p'}\end{equation}
By the same argument that we used in (\ref{then}), 
(\ref{lie4}) is implied by (\ref{lie3}).
Hence, BRST-invariant functionals $W^{-1}$ exist whenever there
are solutions to (\ref{lie1})--(\ref{lie3}), and these solutions
provide all BRST-invariant local functionals for
ghost number $-1$ up to locally trivial ones.
We call equations (\ref{lie1})--(\ref{lie3}) generalized
Killing vector equations because
(\ref{lie1}) reduces in the case $D_a=constant$
to the familiar Killing vector equations,
supplemented by the corresponding condition
(\ref{lie2}) discussed in \cite{Hull,paper1}. 

This implies the results on the rigid symmetries of the models
under study announced and discussed in \cite{letter}.
In particular (\ref{W-1}) yields the infinitesimal
rigid symmetry transformations $\Delta$
of the fields which leave the classical action
invariant. They are just the coefficient functions of the pieces
linear in the antifields of the classical fields:
\begin{eqnarray}
& \Delta \gamma_{\mu\nu}=0,\quad
\Delta X^M= h^M(X)&
\nonumber\\
& \Delta A^a_\mu=
\frac{1}{2}b^a_M(X)\6_\mu X^M-
\frac{1}{2}\sqrt{\gamma}\, \epsilon_{\mu\nu}a^a_M(X)\6^\nu X^M
+A^b_\mu h_b^a(X). &
\label{Delta}\end{eqnarray}
The corresponding Noether currents can be obtained 
from the antifield independent part of the
1-form contained in (\ref{om6}). One finds
\begin{eqnarray}
j^\mu &=&
\sqrt \gamma\, \gamma^{\mu\nu}G_{MN}(X)h^M(X)\6_\nu X^N
\nonumber\\
& &+\epsilon^{\mu\nu}\left[\frac 12\, p_M(X)\6_\nu X^M
-A^a_\nu h^M(X)\6_M D_a(X)\right].
\label{j}
\end{eqnarray}
Because of (\ref{shifts1})--(\ref{shifts3}),
the solutions to (\ref{lie1})--(\ref{lie4}) 
are defined only up to redefinitions
\begin{eqnarray}
a^a_M(X) &\rightarrow& a^a_M(X)+2\hat k^{[ab]}(X)\6_M D_b(X)
\nonumber\\
b^a_M(X) &\rightarrow& b^a_M(X)-2\6_M \hat h^a(X)
-2\hat k^{(ab)}(X)\6_M D_b(X)
\nonumber\\
p_M(X) &\rightarrow& p_M(X)-2\6_M \hat f(X)+2\hat h^a(X)\6_M D_a(X).
\label{redef-1}\end{eqnarray}
These redefinitions modify $W^{-1}$ by BRST-exact pieces.
Accordingly, they change
the rigid symmetry transformations only by special gauge transformations
and on-shell trivial symmetries, as we have shown
explicitly in \cite{letter} to which we also refer
for a more detailed discussion 
of the rigid symmetries. There 
are models with infinitely many inequivalent
rigid symmetries, see \cite{letter} and section \ref{example}.

\subsection{{\boldmath $g=0$}: Gauge invariant actions
and continuous consistent deformations}\label{defs}

Continuous consistent deformations of a gauge invariant action are
modifications of the action and its gauge symmetries,
parametrized by deformation parameters, such
that the original theory is recovered for vanishing deformation parameters
and the deformed action is gauge invariant under the deformed
gauge transformations. They are called trivial if they can be removed
through local field redefinitions. To first order
in the deformation parameters, such deformations are determined by
the local BRST cohomology at ghost number 0 \cite{bh}. It should be 
mentioned however that the existence of such deformations might get
obstructed at higher orders in
the deformation parameters through the BRST cohomology at ghost
number 1. Therefore, in general only a subset of nontrivial solutions to 
(\ref{i1}) with ghost number 0 will eventually yield consistent
deformations. For this reason we shall speak about possible
deformations here.

One may distinguish two types of consistent deformations.
First, there are those deformations which do not change
the gauge transformations. They add
terms to the action which are invariant under the
original gauge transformations. 
In our case (closed gauge algebra) these deformations
arise from the antifield independent solutions to (\ref{i1})
at ghost number 0, and therefore reproduce
the results derived in section
\ref{action}. The deformations of the second type change
both the action and the gauge transformations nontrivially. They arise
from solutions to (\ref{i1}) with
ghost number 0 which depend
nontrivially on antifields. The antifield dependent terms
in these solutions provide the possible deformations
of the gauge transformations and of their algebra to first
order in the deformation parameters.

All inequivalent deformations of the first type arise
from (\ref{om1}) and (\ref{om2}). They
read respectively
\begin{eqnarray}
W^0_1&=&\int d^2\sigma\, \frac 12\, 
        \epsilon^{\mu\nu}h_a(X)F^a_{\mu\nu}
\label{W01}\\
W^0_2&=&\int d^2\sigma\, \frac 12\,
(\sqrt \gamma\, \gamma^{\mu\nu}
+\epsilon^{\mu\nu})\, k_{MN}(X)\,\6_\mu X^M\cdot\6_\nu X^N.
\label{W02}
\end{eqnarray}
These functionals provide deformations which
shift $G_{MN}$, $B_{MN}$ and $D_a$ by arbitrary functions
$k_{(MN)}$, $k_{[MN]}$ and $h_a$ respectively. This
reproduces the result of section \ref{action}.
It should be noted that
some shifts of this type are still cohomologically trivial,
as follows from (\ref{shifts4}) and (\ref{shifts6}).
These shifts are generated by mere local
field redefinitions
\begin{eqnarray*}
& X^M\rightarrow X^M+\hat h^M(X) &
\\
& A^a_\mu \rightarrow A^a_\mu+
\frac{1}{2}\hat b^a_M(X)\6_\mu X^M-
\frac{1}{2}\sqrt{\gamma}\, \epsilon_{\mu\nu}\hat a^a_M(X)\6^\nu X^M
+A^b_\mu \hat h_{0b}^a\ , &
\end{eqnarray*}
where $\hat h_{0b}^a$ is the constant part of $\hat h_b^a$. 
Only this part enters here because the nonconstant
part would introduce antifield dependent terms (it 
contributes in (\ref{shifts3}) and thus in the redefinition
of $b_{b M}^a$ which appears
in the antifield dependent solution $W^0_6$ given below).

The possible deformations of the second type
arise from the total forms (\ref{om6})
and (\ref{om7}) with total degree 2.  (\ref{om6}) yields
the following solutions with ghost number 0,
\begin{eqnarray}
W^0_6 = \int d^2\sigma 
      \left\{-A_\mu^a \, j^\mu_a
      -\frac 12 \, \epsilon^{\mu\nu} A^a_\mu A^b_\nu 
          h^M_a(X) \6_M D_b(X)
      + X^*_MC^{a}h^M_{a}(X)\right. 
\nonumber\\
      + A^{\mu*}_{a}C^{b}\left[\frac 12 b^{a}_{b M}(X)\6_{\mu}X^M 
      - \frac 12\sqrt{\gamma}\, \epsilon_{\mu\nu}
               a^{a}_{b M}(X)\6^{\nu}X^M 
      + A^{c}_{\mu} h^{a}_{cb}(X)\right]
\nonumber \\
\left.
      + \frac 12\, C^*_{a}C^{b}C^{c}h^{a}_{bc}(X)
      + \frac 14\,\epsilon_{\mu\nu}A^{*\nu}_{a}A^{\mu*}_{b}
          C^{c}C^d k^{(ab)}_{cd}(X)\right\}, 
\label{W06} 
\end{eqnarray}
where
\begin{equation}
j^\mu_a =
\sqrt \gamma\, G_{MN}(X)h^M_a(X)\6^\mu X^N
+\epsilon^{\mu\nu}\left[\frac 12\, p_{Ma}(X)\6_\nu X^M
-h^M_a(X) A^b_\nu \6_M D_b(X)\right].
\label{ja}\end{equation}
In (\ref{W06}) we have used already that (\ref{cond6}) gives
\[ f_{ab}(X)=h^M_{[a}(X)\6_M D_{b]}(X). \]
In addition (\ref{cond6}) requires
the target space functions appearing in $W^0_6$ to satisfy
\begin{eqnarray}
 0 &=& h^M_{(a}(X)\6_M D_{b)}(X) 
\label{lie00}\\
 {\cal L}_a G_{MN} &=& -a_{a(M}^b \6_{N)} D_b 
\label{lie01}\\
 {\cal L}_a B_{MN} &=& -\6_{[M} p'_{N]a}-b_{a[M}^b \6_{N]} D_b 
\label{lie02}\\
 {\cal L}_{[a} \6_M D_{b]} &=& h_{ab}^c \6_M D_c 
\label{lie03}\\
 \6_M h_{ab}^c &=& -k^{(c d)}_{ab} \6_M D_d
\label{lie04}
\end{eqnarray}
where ${\cal L}_a$ denotes the Lie derivative along $h_a^M$, and
\[ p'_{Ma}=p_{Ma}+2B_{MN} h^N_a\ .\]
$W^0_6$ is cohomologically nontrivial unless it can be removed
through redefinitions analogous to (\ref{redef-1}),
\begin{eqnarray}
a^b_{a M} &\rightarrow& a^b_{a M}
+2\hat k^{[bc]}_a\6_M D_c
\nonumber\\
b^b_{a M} &\rightarrow& b^b_{a M}-2\6_M \hat h^b_a
-2\hat k^{(bc)}_a\6_M D_c
\nonumber\\
p_{M a} &\rightarrow& p_{M a}
-2\6_M \hat f_a +2\hat h^b_a\6_M D_b\ .
\label{redef0}\end{eqnarray}
Every nontrivial solution to (\ref{lie00})--(\ref{lie04}) solves also
(\ref{lie1})--(\ref{lie4}) nontrivially because
(\ref{lie00}) and (\ref{lie03}) imply
${\cal L}_{a} \6_M D_{b} = h_{ab}^c \6_M D_c$.
The converse is not true, i.e.\ in general (\ref{lie1})--(\ref{lie4})
has nontrivial solutions which do not solve (\ref{lie00})--(\ref{lie04})
and thus do not give rise to solutions $W^0_6$.
Hence, as (\ref{lie1})--(\ref{lie4})
determine the rigid symmetries, the deformations
(\ref{W06}) contain a {\em subset} of rigid symmetries.
This subset can be described as follows.
Labelling the elements of a basis for all nontrivial
rigid symmetries by an index $\alpha$,
the solutions to (\ref{lie1})--(\ref{lie4}) are
sets of target space functions 
\begin{equation} \{h^M_\alpha\, ,\, 
a_{M\alpha}^a\, ,\, b_{M\alpha}^a\, ,\, p_{M \alpha}\, ,\, 
h_{a \alpha}^b\, ,\, k_{a \alpha}^{(bc)}\}.
\label{rigidbasis}\end{equation}
The solutions to (\ref{lie00})--(\ref{lie04}) are then
linear combinations of the functions (\ref{rigidbasis})
with constant coefficients $\lambda^\alpha_a$,
\begin{equation} 
h^M_a=\lambda^\alpha_a\, h^M_\alpha\ ,\ \dots\ ,\ 
k^{(c d)}_{ab}=\lambda^\alpha_{[a}\, k^{(c d)}_{b]\alpha}\ ,
\end{equation}
subject to the following requirement imposed by (\ref{lie00}),
\begin{equation}
h^M_\alpha\, \6_M  D_{(a} \lambda^\alpha_{b)}=0.
\label{condition}\end{equation}
Accordingly, one has
\begin{equation} 
j^\mu_a =\lambda^\alpha_a\, j^\mu_\alpha
\label{currentLK}\end{equation}
where $j^\mu_\alpha$ is the Noether current (\ref{j})
which corresponds to the $\alpha$th rigid symmetry.

We conclude that
deformations arising from $W^0_6$ gauge 
rigid symmetries satisfying (\ref{condition}) and deform
the abelian gauge transformations corresponding
to the $A_\mu^a$.
The presence of the term in $W^0_6$ 
involving $C^*_a$ indicates that
the algebra of
the deformed gauge transformations will not be abelian anymore
whenever $h^a_{bc}\neq 0$. The
term quadratic in the $A^{\mu*}_a$ shows that the algebra of
the deformed gauge transformations closes off-shell only
if $k^{(ab)}_{cd}= 0$.

Some deformations arising from $W^0_6$ are obvious, namely
standard nonabelian extensions of Lagrangians (\ref{L})
where the abelian field strengths turn into nonabelian ones
and the matter fields are coupled minimally to the
gauge fields (through standard
covariant derivatives). Among them there are deformations of
particular models which, after elimination of auxiliary fields, 
reproduce nonabelian Born-Infeld type
Lagrangians \cite{nonabBI},
\begin{equation}
L^{nonabel.}_{BI}=
\mbox{Tr} \sqrt{-\det(\mbox{{\bf 1}}{\cal G}_{\mu\nu}
+\mbox{{\bf F}}_{\mu\nu})}\ .
\label{nonabelian}
\end{equation}
Here $\mbox{{\bf F}}_{\mu\nu}$ is a Lie algebra 
valued matrix containing the
nonabelian field strengths, {\bf 1} is the unit matrix,
and ${\cal G}_{\mu\nu}$ is an induced
world-sheet metric as in (\ref{BI4}). 

Details and more general deformations of this type 
are given in appendix \ref{appC}.

(\ref{om7}) yields the following solutions with ghost number 0,
\begin{eqnarray}
W_{7\pm}^0 = \int d^2\sigma \left\{
    - 2\6_{\pm}h_{\mp\mp}\cdot h^{M\pm}(X)G_{MN}(X){\cal D}_{\pm}X^N
    +X^*_M\, \Theta_{\pm}h^{M\pm}(X)\right.
\nonumber\\
\left. +\frac 12\,A^{\mu*}_{a}\, \Theta_{\pm}
       [b^{a\pm}_M(X)\6_{\mu}X^M
       -\sqrt{\gamma}\, \epsilon_{\mu\nu} a^{a\pm}_M(X)\6^{\nu}X^M]
\right\}
\label{W07}
\end{eqnarray}
with $h_{\mp\mp}$ and ${\cal D}_{\pm}X^M$ as in (\ref{red1}) and (\ref{Ts})
respectively, and 
\begin{equation}
\Theta_\pm=\6_\pm \xi^\pm+h_{\mp\mp}\6_\pm\xi^\mp.
\label{Theta}
\end{equation}
The target space functions $h^{M\pm}(X)$, $a_M^{a\pm}(X)$
and $b_M^{a\pm}(X)$ in $W^{0\pm}_7$ have to
satisfy the partial differential equations (\ref{cond7}). 
$W^{0\pm}_7$ is nontrivial
unless it can be removed through the cohomologically
trivial redefinitions
\begin{eqnarray}
a^{a\pm}_{M}(X) &\rightarrow& a^{a\pm}_{M} (X)
                + 2\hat{k}^{[ab]\pm}(X)\,\6_M D_{b}(X)
\nonumber\\
b^{a\pm}_{M}(X) &\rightarrow& b^{a\pm}_{M} (X)
- 2\hat{k}^{(ab)\pm}(X)\,\6_M D_{b}(X).
\label{n4} 
\end{eqnarray}
Every solution to (\ref{cond7}) is also a solution to
(\ref{lie1})--(\ref{lie4}) with $h_a^b=k_a^{(bc)}=0$.
Furthermore, such a solution to (\ref{lie1})--(\ref{lie4})
is nontrivial whenever the solution to (\ref{cond7}) is
nontrivial too\footnote{Indeed, assume the solution
to (\ref{lie1})--(\ref{lie4}) arising from a nontrivial solution to
(\ref{cond7}) were trivial. This means
$h^{M\pm}=0$, $a^{a\pm}_{M}=-2\hat k^{[ab]}\6_M D_b$,
$b^{a\pm}_{M}=2\6_M\hat h^a+2\hat k^{(ab)}\6_M D_b$
and $p^\pm_M=2\6_M\hat f-2\hat h^a\6_M D_a$.
Since solutions to (\ref{cond7}) fulfill
$p^\pm_M=\mp 2 G_{MN}h^{N\pm}$, one gets
$\6_M\hat f=\hat h^a\6_M D_a$ which implies
$\6_M \hat h^a=\Lambda^{(ab)}\6_M D_b$ for some
$\Lambda^{(ab)}$ by the same arguments used to derive
(\ref{then}). Altogether this contradicts
that the solution to (\ref{cond7}) is nontrivial,
because we get $h^{M\pm}=0$, $a^{a\pm}_{M}=-2\hat k^{[ab]}\6_M D_b$
and $b^{a\pm}_{M}=2\hat k'{}^{(ab)}\6_M D_b$ with
$\hat k'{}^{(ab)}=\hat k^{(ab)}+\Lambda^{(ab)}$.}.
Hence, the nontrivial solutions to (\ref{cond7}) are also a subset
of rigid symmetries.
The antifield independent term in $W^{0+}_7$ ($W^{0-}_7$) contains
the $-$ ($+$) component of the Noether currents (\ref{j}) corresponding to
these special rigid symmetries due to
$2 h^{M\pm}G_{MN}{\cal D}_{\pm}X^N = j^\mp$.

As $W^{0\pm}_7$ involves the diffeomorphism ghosts, 
corresponding deformations modify the world-sheet diffeomorphisms. 
The algebra of the diffeomorphisms will however
not get deformed at first order in the deformation parameters
because $W^{0\pm}_7$ contains neither antifields 
of ghosts nor terms quadratic in antifields.
Analogous deformations of the world-sheet diffeomorphisms 
in standard sigma models have been found and discussed 
in \cite{paper2}.
\medskip

\noindent {\bf Remark:}

The existence of a cocycle (\ref{W06}) implies that
there are infinitely many rigid symmetries. In other words,
an action with Lagrangian (\ref{L}) has infinitely many 
rigid symmetries whenever the set of equations 
(\ref{lie00})--(\ref{lie04}) has a nontrivial solution.
Namely, one easily verifies that 
the antibracket $(W^{-2},W^0_6)$ 
of cocycles (\ref{W-2}) and (\ref{W06}) has precisely the
form (\ref{W-1}) with $h^M(X)=h^a(X)h_a^M(X)$ etc..
Now, the antibracket of two BRST cocycles
is again a BRST cocycle. Furthermore, a cocycle (\ref{W-1}) 
corresponds to a nontrivial rigid symmetry whenever
$h^M(X)\neq 0$. As there are always infinitely many
cocycles $W^{-2}$ (see section \ref{2ndorder}), we conclude
that any cocycle $W^0_6$ gives indeed rise to infinitely
many rigid symmetries via the
antibrackets $(W^{-2},W^0_6)$.

\subsection{{\boldmath $g=1$}: Candidate gauge anomalies}

The candidate gauge anomalies arise from (\ref{om1}), (\ref{om2})
and (\ref{om4})--(\ref{om7}). Those obtained from (\ref{om1}) and 
(\ref{om2}) can be subdivided into two subsets involving the
abelian ghosts $C^a$ and the diffeomorphism ghosts respectively,
\begin{eqnarray}
W^1_{1\pm} &=& \int d^2\sigma\, \frac {1}{2}\,\Theta_{\pm}
               \epsilon^{\mu\nu}F^{a}_{\mu\nu}h_a^\pm(X)
\label{W11pm}\\
W^1_{2\pm} &=& \int d^2\sigma\, \frac {1}{2}\,\Theta_{\pm}
               (\sqrt{\gamma}\gamma^{\mu\nu}+ \epsilon^{\mu\nu})\,
               k^{\pm}_{MN}(X)\,\6_{\mu}X^M\cdot\6_{\nu}X^N
\label{W12pm}\\
W^1_{1C} &=& \int d^2\sigma \,\frac {1}{2}\, C^{a}
     \epsilon^{\mu\nu}F^{b}_{\mu\nu}h_{ab}(X)
\label{W11C}\\
W^1_{2C} &=& \int d^2\sigma \,\frac {1}{2}\, C^{a}
             (\sqrt{\gamma}\gamma^{\mu\nu}+ \epsilon^{\mu\nu})\,
             k_{MNa}(X)\,\6_{\mu}X^M\cdot\6_{\nu}X^N
\label{W12C} 
\end{eqnarray}
with $\Theta_\pm$ as in (\ref{Theta}). These functionals
are BRST invariant for any choice of the target space functions 
occurring in them. They are nontrivial except for those which
can be removed through redefinitions obtained from
(\ref{shifts4}) and (\ref{shifts6}).

Note that the candidate anomalies (\ref{W11pm}) and (\ref{W12pm})
come in two chiralities. The solutions (\ref{W12pm}) are
familiar from ordinary sigma models where particular
(``left-right symmetric'') linear combinations of such solutions 
provide the conformal anomalies involving the
$\beta$-functions corresponding to $G_{MN}$ and $B_{MN}$,
see \cite{paper1} for details. Therefore one would expect
(\ref{W11pm}) and (\ref{W12pm}) to have a similar interpretation
in the models under study.

The candidate  anomalies (\ref{W11C}) and (\ref{W12C}) 
are of the form ``abelian ghost 
$\times$ Weyl- and $U(1)$-invariant scalar density''. 
Such solutions are
familiar from Yang--Mills theory and gravity \cite{comgrav}. 
In particular, (\ref{W11C}) reproduces
the representatives of standard abelian chiral anomalies 
for the special choice $h_{ab}=h_{(ab)}=constant$. 

(\ref{om6}) and (\ref{om7}) yield
the following candidate gauge anomalies,
\begin{eqnarray}
 W^1_6 &=&  \int d^2\sigma  \left\{
- C^a A_\mu^b \, 
[\sqrt \gamma\, G_{MN}(X) h^M_{ab}(X) \6^\mu X^N
+\frac 12\epsilon^{\mu\nu}p_{M ab}(X)\6_\nu X^M]
\right.
\nonumber\\
& &
-\frac 12 C^{[a} A_\mu^b A_\nu^{c]} 
\epsilon^{\mu\nu} h^M_{ab}(X)\,\6_M D_{c}(X)
+ \frac {1}{2}X^*_M C^{a}C^{b}h^M_{ab}(X)
\nonumber\\
& &
+\frac {1}{4}A^{\mu*}_{a}C^b C^c
[b^{a}_{Mbc}(X)\6_{\mu}X^M
-\sqrt{\gamma}\, \epsilon_{\mu\nu}a^{a}_{Mbc}(X)\6^{\nu}X^M
+2A^{d}_{\mu} h^{a}_{bcd}(X)]
\nonumber \\
& & 
\left.
+\frac {1}{6}C^{a}C^{b}C^{c}
[C^*_{d}h^{d}_{abc}(X)
+\frac {1}{2}\epsilon_{\mu\nu}A^{*\nu}_{e}A^{\mu*}_{d}
k^{(ed)}_{abc}(X)]
\right\} 
\label{W16} \\
 W_{7\pm}^1 &=&  \int d^2\sigma \left\{
2[\Theta_{\pm}
(A^{a}_{\mp} - h_{\mp\mp}A^{a}_{\pm})
-C^{a}\6_{\pm} h_{\mp\mp}\cdot]
G_{MN}(X)h^{M\pm}_{a}(X){\cal D}_{\pm}X^N
\right. 
\nonumber\\
& &
-\frac 12 C^{a}\Theta_{\pm}A^{\mu*}_{b}
[b_{Ma}^{b\pm}(X) \6_{\mu}X^M
-\sqrt \gamma\epsilon_{\mu\nu} a_{Ma}^{b\pm}(X) \6^{\nu}X^M]
\nonumber\\
& &
\left.
-C^{a}\Theta_{\pm}X^*_M h^{M\pm}_{a}(X)
\right\},
\label{W17}
\end{eqnarray}
where the target space functions have to satisfy 
sets of partial differential equations analogous
to (\ref{lie00})--(\ref{lie04}) and (\ref{cond7})
respectively. These solutions are nontrivial except for
those we can be removed through redefinitions analogous to
(\ref{redef-1}) and (\ref{n4}).
Both (\ref{W16}) and (\ref{W17}) thus involve 
subsets of rigid symmetries.
The antifield independent parts of (\ref{W16}) and (\ref{W17})
involve the corresponding Noether currents.

Note that
solutions (\ref{W16}) occur only if there are at least two abelian
gauge fields. In this respect they are 
somewhat similar to by now well-known
candidate anomalies in Yang--Mills theory and gravity
\cite{2U1s}. (\ref{W17})
gives candidate gauge anomalies which involve
necessarily both
the abelian ghosts $C^a$ and the diffeomorphism ghosts. These
solutions might therefore have an interpretation as mixed anomalies for
world-sheet diffeomorphisms and $U(1)$-transformations.

Finally, (\ref{om5}) gives the following candidate anomalies
\begin{eqnarray}
W_{8\pm}^1 = \pm \int d^2\sigma 
\left\{ 
g^{\pm\pm}(X)[\Theta_{\pm} \6^2_{\pm} h_{\mp\mp}
-\6_{\pm} h_{\mp\mp}\cdot(\6_{\pm}\Theta_{\pm}
+\6_{\pm} h_{\mp\mp}\cdot\6_\pm\xi^{\mp})]
\right.
\nonumber\\
\left.
-(A^{\pm*}_{a} + h_{\mp\mp}A^{\mp*}_{a})k^{a \pm}_{\pm}(X)
\Theta_{\pm}(\6_{\pm}^2\xi^{\pm}+h_{\mp\mp}\6_{\pm}^2\xi^{\mp}
- 2 \6_{\pm}h_{\mp\mp}\cdot\6_{\pm}\xi^{\mp}) \right\}
\label{g33}
\end{eqnarray}
where 
\begin{equation}
\6_M g^{\pm\pm} \mp k^{a \pm}_{\pm}\6_M D_{a} = 0.
\label{gpmpm}\end{equation}
Up to the constant parts of $g^{++} $ and $g^{--}$,
the solutions to (\ref{gpmpm}) determine the
conservation laws of second order, cf.\ section \ref{2ndorder}.
Hence, these conservation laws enter the candidate anomalies 
(\ref{g33}).
The constant parts of $g^{\pm\pm}$ provide
antifield independent solutions arising from (\ref{om5o}).
Fixing these constants to $\mp 1/2$ and
performing a partial integration, the corresponding solutions read
\begin{equation} \stackrel{o}{W}{\!}^1_{8\pm} =\int d^2\sigma \,
(\xi^\pm+h_{\mp\mp}\xi^\pm)\, \6_{\pm}^3h_{\mp\mp}\quad .
\label{familiar}\end{equation}
These solutions are familiar from ordinary sigma
models where their left-right symmetric linear combination
is cohomologically equivalent to the conformal anomaly 
$\int d^2\sigma c\,\sqrt \gamma\, R$ \cite{paper1}.

\mysection{Example} \label{example}

In order to illustrate the general results we shall now specify
them for a simple class of models which we treated already in \cite{letter}.
These models contain only one abelian gauge field $A_\mu$ 
and besides are characterized by
\begin{eqnarray}
& G_{0M}=0,\ G_{mn}=g(\varphi)\,\eta_{mn} &
\nonumber\\
& B_{0m}=0,\ B_{mn}=B_{mn}(\varphi),\ D=D(\varphi), &
\label{simple}
\end{eqnarray}
where we used the same notation as in (\ref{BI1}). We will
carry out the analysis for any choice of $g(\varphi)$, 
$B_{mn}(\varphi)$ and
$D(\varphi)$ except that we assume $D\neq constant$ (as the last
term in (\ref{L}) becomes a total derivative for $D=constant$).

Since these models contain only one abelian gauge field,
nontrivial solutions to (\ref{i1}) exist only
at ghost numbers ranging from $-2$ to 5 (in general,
section \ref{Result} provides
solutions at ghost numbers $-2,\dots,4+N$ where $N$ is the number of
abelian gauge fields). As the specification of the antifield
independent solutions is straightforward, we shall only
discuss the antifield dependent ones.
This amounts to solving the partial differential equations
(\ref{cond6})--(\ref{cond5}) in the various ghost number sectors.
	
\paragraph{{\boldmath $g=-2$}: Second order conservation laws.}

The solutions to (\ref{secsymm}) are
in this case $h^a\equiv h(\varphi)$ and 
$k^{(ab)}\equiv -h'(\varphi)/D'(\varphi)$,
where $h(\varphi)$ is an arbitrary function of $\varphi$ and
the prime denotes differentiation with respect to $\varphi$,
\[ '=\frac{\6}{\6\varphi}\ . \]
The nontrivial solutions of (\ref{i1}) at ghost number $-2$
are thus
\begin{equation}
W^{-2}=\int d^2\sigma\left[C^* h(\varphi)+
\frac 12\, \epsilon_{\mu\nu}
A^{\mu *} A^{\nu *}\frac{h^{'}(\varphi)}{D^{'}(\varphi)} \right].
\label{W-2a}\end{equation}
The corresponding second order conservation laws are
functions $f(\varphi)$ related to
$h(\varphi)$ by (\ref{secsymm2}),
i.e.\ one has $h= f'/ D'$. 

\paragraph{{\boldmath $g=-1$}: Rigid symmetries and Noether currents.}

(\ref{lie1})--(\ref{lie3}) have been solved already in \cite{letter}.
(\ref{lie4}) just determines $k\equiv k^{(bc)}_a$ in terms of
$h\equiv h^b_a$ and $D$. Altogether, the general solution
of (\ref{lie1})--(\ref{lie4}) reads, up to redefinitions
(\ref{redef-1}),
\begin{eqnarray}
h^0&=&h^0(\varphi)
\nonumber\\
h^m&=&- h^0(\varphi) x^m g'(\varphi)/[2g(\varphi)]
+r^m(\varphi)+r^{[mn]}(\varphi)\eta_{nk}x^k
\nonumber\\
a_m&=&-2\eta_{mn} h^n{}'g/D',\quad
a_0= 0
\nonumber\\
b_m&=&[2B'_{mn}h^n+(B'_{mn}h^0)'x^n]/D',\quad
b_0=0
\nonumber\\
p_m&=&B'_{mn}h^0x^n,\quad  p_0 =0
\nonumber\\
h&=&h^0{}' + h^0 D''/D'
\nonumber\\
k&=& -h'/D'.
\label{sol}
\end{eqnarray}
Here $h^0$, $r^m$ and $r^{[mn]}$ are arbitrary functions
of $\varphi$.
The corresponding solutions to (\ref{i1}),
rigid symmetries and Noether currents are
obtained by inserting (\ref{sol}) in (\ref{W-1}),
(\ref{Delta}) and (\ref{j}) respectively. As discussed in
\cite{letter}, these infinitely many rigid symmetries
constitute a loop version of the Weyl algebra (= Poincar\'e
algebra + dilatations). Thereby $h^0$, $r^m$ and $r^{[mn]}$
give rise to generalized dilatations, translations and
Lorentz transformations respectively.

\paragraph{{\boldmath $g=0$}: 
Deformations of the gauge symmetries.}

The antifield dependent solutions of (\ref{i1}) with ghost number 0
are determined by Eqs.\ (\ref{lie00})--(\ref{lie04}) and (\ref{cond7})
respectively.
As explained in section \ref{defs}, the nontrivial solutions to
(\ref{lie00})--(\ref{lie04})
form a subset of the nontrivial
solutions to (\ref{lie1})--(\ref{lie4}) subject to (\ref{lie00}).
In this case (\ref{lie00}) yields $h^M\6_M D = 0 \Leftrightarrow h^0=0$.
The nontrivial solutions to (\ref{lie00})--(\ref{lie04}) are
thus obtained from (\ref{sol}) simply by setting $h^0=0$ there,
\begin{eqnarray}
h^0&=&0
\nonumber\\
h^m&=&r^m(\varphi)+r^{[mn]}(\varphi)\eta_{nk}x^k
\nonumber\\
a_m&=&-2\eta_{mn} h^n{}'g/D',\quad
a_0= 0
\nonumber\\
b_m&=&2B'_{mn}h^n/D',\quad
b_0=0
\nonumber\\
p_M&=&0,\quad h=0, \quad k=0.
\label{W06a}
\end{eqnarray}
Inserting this in (\ref{W06}), we get
\begin{eqnarray}
W^0_6 = \int d^2\sigma \left\{
      \frac 12 A^{\mu*} C [b_m\6_{\mu}x^m 
      - \sqrt{\gamma}\, \epsilon_{\mu\nu}
               a_m \6^{\nu}x^m ]
\right.
\nonumber \\
\left.
      + x^*_m C h^m(X)
      - \sqrt{\gamma}\, A_{\mu} g(\varphi)\eta_{mn}h^m \6^\mu x^n
\right\}
\label{W06c} 
\end{eqnarray}
with $h^m$, $a_m$ and $b_m$ as in (\ref{W06a}). Note that
$W^0_6$ contains the generalized translations and Lorentz transformations,
but no dilatations. Hence, corresponding deformations will not
gauge the generalized dilatations.

(\ref{cond7}) requires in this case in particular 
$h^{M\pm}\6_M D = 0$ and $\eta_{nk}\6_m h^{k\pm} = 0$. 
It is then straightforward to verify that the general 
solution to (\ref{cond7}) is, up to trivial redefinitions (\ref{n4}),
\begin{eqnarray}
& h^{0\pm}=0,\quad h^{m\pm}=r^{m\pm}(\varphi)
\nonumber\\
& a^{\pm}_m = -2\eta_{mn}r^{n\pm}{}'g/D',\quad a^{\pm}_0 =0 &
\nonumber\\
& b^{\pm}_m =2[B'_{mn}r^{n\pm} \mp \eta_{mn}(g r^{n\pm})']/D',\quad
b^{\pm}_0 =0, &
\label{barca}
\end{eqnarray}
where $r^{m\pm}(\varphi)$ are arbitrary functions.
Hence, only the generalized translations, but neither the
dilatations nor the Lorentz transformations enter the 
corresponding solutions (\ref{W07}) which read
\begin{eqnarray}
W_7^{0\pm} = \int d^2\sigma \left\{
      \frac 12\,A^{\mu*}\, \Theta_{\pm}
      [b^{\pm}_m(\varphi)\6_{\mu}x^m
       -\sqrt{\gamma}\, \epsilon_{\mu\nu} a^{\pm}_m(\varphi)\6^{\nu}x^m]
\right.
\nonumber\\
\left.
      +x^*_m\, \Theta_{\pm}r^{m\pm}(\varphi)
      - 2\6_{\pm}h_{\mp\mp}\cdot \eta_{mn}r^{m\pm}(\varphi)
      g(\varphi){\cal D}_{\pm}x^n
\right\}.
\label{W07b}
\end{eqnarray}

\paragraph{{\boldmath $g=1$}: Antifield dependent candidate gauge anomalies.}
Solutions with ghost number 1 depending nontrivially on antifields
arise only from (\ref{W17}) and (\ref{g33}), as (\ref{W16})
can give solutions only if there are at least two abelian gauge fields.
(\ref{W17}) gives
\begin{eqnarray}
W_7^{1\pm} = \int d^2\sigma \left\{
 2[\Theta_{\pm}(A_{\mp} -h_{\mp\mp}A_{\pm})-C\6_{\pm}h_{\mp\mp}\cdot ]
\eta_{mn} r^{n\pm}(\varphi) g(\varphi){\cal D}_{\pm}x^m \right.
\nonumber\\
\left.
 +x^*_m C \Theta_{\pm} r^{m\pm}(\varphi)
 +\frac 12\,C\Theta_{\pm}A^{\mu*}\, [b^{\pm}_m(\varphi)\6_{\mu}x^m
 -\sqrt{\gamma}\, \epsilon_{\mu\nu} a^{\pm}_m(\varphi)\6^{\nu}x^m]
\right\}
\label{W17b}
\end{eqnarray}
with $a^{\pm}_m$ and $b_m^{\pm}$ as in (\ref{barca}). 

The solutions to (\ref{gpmpm}) are 
\begin{equation} g^{\pm\pm}=g^{\pm\pm}(\varphi),\quad 
k_\pm^\pm=\pm g^{\pm\pm}{}'(\varphi)/D'(\varphi).
\end{equation}
Inserting this into (\ref{g33}), one obtains the remaining antifield
dependent candidate anomalies.

\paragraph{{\boldmath $g>1$}.}
Further antifield dependent solutions exist in this case only
at ghost number 2. They arise from (\ref{om5}) through the
$\7C$-dependent contributions to $g^{\pm\pm}$ and $k_\pm^\pm$.

\mysection{Conclusions} \label{conclusions}

Let us finally summarize our main results for
models with the field content, gauge symmetries and locality
requirements described in section \ref{intro} and \ref{setup}. 
First,
we have {\em proved} the results announced and discussed already
in \cite{letter}, that is:

a) Any Lagrangian describing such a model can be cast in the form 
(\ref{L}). The class of models can thus be parametrized 
through target space functions 
$G_{MN}(X)$, $B_{MN}(X)$ and $D_a(X)$ where $G_{MN}$ and $B_{MN}$
are symmetric and antisymmetric respectively. Particular 
choices of these functions yield Born--Infeld actions upon elimination
of auxiliary fields. We note that none of our results singles out
Born--Infeld actions in the larger class of models studied here.
Of course, this does not exclude the existence of other criteria which
could distinguish them, such as T duality or the existence of 
supersymmetric extensions with local kappa symmetry.

b) The nontrivial rigid symmetries and Noether currents
of such models are precisely given by
(\ref{Delta}) and (\ref{j}) in terms of the solutions to
the generalized Killing vector equations
(\ref{lie1})--(\ref{lie4}). These equations can have infinitely many
inequivalent solutions, i.e.\ there are models with infinitely many
nontrivial rigid symmetries and Noether currents.

In addition we have derived the following results:

c) Apart from the Noether currents, there are infinitely many
dynamical conservation laws of second order and corresponding
second order symmetries of the master equation, 
see section \ref{2ndorder}. The occurrence of an infinite number
of such conservation laws originates from the fact that the
world-sheet is two dimensional and is familiar from
two dimensional Maxwell theory, see \cite{bbh1}, section 12. 
We have also seen that this implies an infinite number of
rigid symmetries of the classical action whenever there is at least one
nontrivial solution to equations (\ref{lie00})--(\ref{lie04}),
see remark at the end of section \ref{defs}.

d) The possible nontrivial 
consistent deformations of the gauge transformations arise from
solutions to Eqs.\ (\ref{lie00})--(\ref{lie04}) and (\ref{cond7}).
These equations determine subsets of the rigid symmetries respectively.
The consistent deformations thus gauge some of the rigid symmetries.
The accordingly deformed actions involve the corresponding
Noether currents. Deformations arising from
solutions to (\ref{lie00})--(\ref{lie04}) deform the
abelian gauge transformations. Particular examples for such
deformations give rise to standard nonabelian Born--Infeld actions.
Other deformations of this kind might even lead to an open gauge algebra.
Deformations involving solutions to (\ref{cond7}) deform
the world-sheet diffeomorphisms nontrivially. 
It could be interesting to examine whether such deformations can provide
manifestly $SL(2,Z)$ invariant formulations of the D-string as in
\cite{paul,cederwall}.

e) There are three types of antifield independent candidate anomalies.
Those of the first type, given by (\ref{W11C}) and (\ref{W12C}),
involve the abelian ghost fields, those of
the second type, given by (\ref{W11pm}) and (\ref{W12pm}), 
the diffeomorphism ghosts. The former take the
form ``abelian ghost $\times$ density of a gauge invariant action'',
the latter take an analogous form where the abelian ghosts 
are replaced by (\ref{Theta}). The third type
of antifield independent candidate anomalies contains the two
solutions (\ref{familiar}) which are familiar from ordinary
sigma models. 

f) In sharp contrast to the result on potential
anomalies of ordinary sigma models \cite{paper1}, 
there may also be candidate anomalies from which the
antifield dependence {\em cannot} be removed.
One may distinguish three types of them.
Those of the first type, given by (\ref{W16}), 
involve a subset of the rigid symmetries
determined by equations analogous to (\ref{lie00})--(\ref{lie04}).
They contain the abelian ghosts, but not the diffeomorphism ghosts,
and can occur only in presence of at least two abelian gauge fields.
Those of the second type, given by (\ref{W17}), 
involve the subset of the rigid symmetries
determined by (\ref{cond7}) and contain both the abelian and the
diffeomorphism ghosts. Those of the third type, given by
(\ref{g33}), involve the second order conservation laws
and contain only the diffeomorphism ghosts.
To our knowledge, none of these antifield dependent candidate 
anomalies was previously known for models studied here.

\section*{Acknowledgements}

JS thanks Fundaci\'o Agust\'{\i} Pedro i Pons 
for financial support.
FB was supported by the Spanish ministry of education and 
science (MEC). Furthermore
this work has been partially supported by AEN95-0590 (CICYT),
GRQ93-1047 (CIRIT) and by the 
Commission of European Communities CHRX93-0362 (04).

\appendix

\mysection{Spurious gauge symmetries}\label{appA}
	
In this appendix we discuss some conditions which have to be imposed on
the Lagrangians (\ref{L}) in order to prevent the occurrence of
additional gauge symmetries. We shall also argue that these conditions
are natural as they just eliminate spurious gauge symmetries.

Assume one would choose $D_a(X)$ such that there are target space 
functions $k^a(X)$ satisfying $k^a\6_M D_a=0$. Then the
action would be invariant (up to a surface term) under 
\[
\delta_\varepsilon \gamma_{\mu\nu}=\delta_\varepsilon X^M=0,\quad
\delta_\varepsilon A^a_\mu=\varepsilon_\mu k^a(X)
\]
where the $\varepsilon_\mu$
are arbitrary fields. Hence, in such a case the action would have
extra gauge symmetries and one would have to introduce
extra ghost fields $C_\mu$ and their antifields $C^{\mu *}$.
Therefore we require in particular that the functions $D_a(X)$ are
such that
\begin{equation}
   k^a\6_M D_a=0 \quad \Rightarrow \quad k^a=0.
\label{extra}
\end{equation}
This condition appears natural. Indeed,
assume for instance that each function $D_a$ depends only on
one target space coordinate and that all these coordinates are
different (in fact, this can be often achieved by
an appropriate target space reparametrization).
Then (\ref{extra}) is always satisfied, provided
none of the $D_a$ is constant. As a constant $D_a$ yields
only a contribution to the Lagrangian which is a total derivative,
the corresponding gauge field $A_\mu^a$ actually does not
contribute to the classical action, and
we can indeed assume (\ref{extra}) with no loss of
generality in such a case.

More generally we require
\begin{eqnarray}
& G_{MN}\,\eta^N=\lambda^a \6_M D_a\ ,\quad \eta^M\6_M D_a=0 &
\label{extra1a}\\
& \6_{(M}[\lambda^a \6_{N)} D_a]-\Gamma_{MNK}\,\eta^K
+\frac 12 a_{(M}^a\6_{N)}D_a=0 &
\\
& H_{MNK}\,\eta^K+b_{[M}^a\6_{N]}D_a=0 &
\label{extra1c}\\
\Rightarrow 
&
\eta^M=\lambda^a=0,\ \exists\Lambda^{ab}:\ 
a_M^a=\Lambda^{[ab]}\6_M D_b\ ,\ 
b_M^a=\Lambda^{(ab)}\6_M D_b.
&
\label{extra2}
\end{eqnarray}
Namely, assume there were functions $\eta^M(X)$, $\lambda^a(X)$,
$a_M^a(X)$ and $b_M^a(X)$ satisfying Eqs. (\ref{extra1a})--(\ref{extra1c}).
Then the
Lagrangian would be invariant up to a total derivative under
\begin{eqnarray}
& \delta_\varepsilon \gamma_{\mu\nu}=0,\quad
\delta_\varepsilon X^M=\varepsilon\, \eta^M(X) &
\nonumber\\
& \delta_\varepsilon A_\mu^a=
\sqrt{\gamma}\, \epsilon_{\mu\nu}\lambda^a(X) \6^\nu \varepsilon
+\frac 12\,\varepsilon\,[b_M^a(X)\6_\mu X^M
-\sqrt{\gamma}\,\epsilon_{\mu\nu} a_M^a(X)\6^\nu X^M] &
\label{spur}\end{eqnarray}
where $\varepsilon$ is an arbitrary field.
This would mean that the action has an additional gauge
symmetry unless (\ref{extra2}) holds (for (\ref{extra2}) implies that
the transformations $\delta_\varepsilon$ are on-shell trivial and 
thus do not establish additional gauge symmetries).
Such gauge symmetries appear to be spurious too.
This is suggested by the following
toy model which has only one abelian gauge field
$A_\mu$ and besides is defined through
\[
G_{MN}=\left( \begin{array}{c|c}
\begin{array}{cc}
0 & 1 \\ 1 & 0
\end{array} & 0\\
\hline
0 & \mbox{{\bf 1}}
\end{array}
\right),\ B_{MN}=0,\ D = X^1\quad (M,N=0,1,\dots).
\]
It is easy to see that, for this model,
the general solution to equations
(\ref{extra1a})--(\ref{extra1c}) is, modulo
solutions of the form (\ref{extra2}),
\[
\eta^M = \delta^M_0\lambda(X), \quad 
 a_M = -2\6_M\lambda(X),\quad 
b_M=0
\]
where $\lambda(X)$ is a completely arbitrary
target space function. This reflects that the action
is gauge invariant under
\[
\delta_{\bar \varepsilon}\gamma_{\mu\nu}=0,\quad
\delta_{\bar \varepsilon} X^M={\bar \varepsilon}\, \delta^M_0,\quad
\delta_{\bar \varepsilon} A_\mu=
\sqrt{\gamma}\,\epsilon_{\mu\nu}\6^\nu {\bar \varepsilon}.
\]
[From this one recovers (\ref{spur}) for
${\bar \varepsilon}=\varepsilon\lambda$.]
As $\delta_{\bar \varepsilon}$ generates the shift symmetry 
$X^0\rightarrow X^0+{\bar \varepsilon}$, it is not surprising that
the Lagrangian can be cast in a form such that $X^0$ drops out.
Indeed, up to a total derivative the Lagrangian reads
\[
L_0 =
\sum_{M>1}\frac 12\,\sqrt{\gamma}\,\6_\mu X^M \6^\mu X^M
-\epsilon^{\mu\nu}\6_{\mu}X^1\cdot (A_{\nu}
-\sqrt{\gamma}\epsilon_{\nu\rho}\6^\rho X^0),
\]
which shows that $X^0$ can be removed through
the field redefinition
\[
A'_\mu = A_{\mu}-\sqrt{\gamma}\,\epsilon_{\mu\nu}\6^\nu X^0.
\]
Hence, the gauge symmetry
(\ref{spur}) itself can in this case be removed through 
a mere local field redefinition and is thus indeed spurious.
	
\mysection{Second part of the cohomological analysis}
\label{appB}

In this appendix we derive the results described in section \ref{Result} 
by applying $\7s$ to (\ref{om}) and analysing $\7s\omega=0$ modulo
$\omega\rightarrow\omega-\7s\hat\omega$ where $\hat\omega$ has
the same form as $\omega$.
We will use the notation
\begin{eqnarray}
\omega &=& {\cal F}(z)+y_\mu^A\, {\cal G}^\mu_A(z)
+y_{+-}^\Delta{\cal H}_\Delta(z)+y_+^Ay_-^B {\cal K}_{AB}(z)
\\
y_\mu^A\, {\cal G}_A^\mu(z) &=&
\7X_+^M{\cal G}_M^+(z) + \7Y_+{\cal G}^+(z) + \7A^{- *}_{a +}{\cal G}^{+a}(z)
+ (+\leftrightarrow -)
\label{Gs}\\
y_{+-}^\Delta {\cal H}_\Delta(z) &=& 
\7C^*_{a+-} {\cal H}^a(z)
+\7A^{+ *}_{a +-}{\cal H}_+^a(z)+\7A^{- *}_{a +-}{\cal H}_-^a(z)
\nonumber\\
& & +\7X^*_{M+-}{\cal H}^M(z)+\7X^M_{+-}{\cal H}_M(z)+\7F^a_{+-}{\cal H}_a(z)
\label{Hs}\\
y_+^A y_-^B {\cal K}_{AB}(z) &=&
\7X_+^M \7X_-^N {\cal K}_{MN}(z)+\7X_+^M \7Y_- {\cal K}_{M-}(z)
+\7Y_+ \7X_-^M {\cal K}_{+ M}(z)
\nonumber\\
& & +\7X_+^M \7A_{a-}^{+*} {\cal K}_M{}^a(z)
+\7A_{a+}^{-*} \7X_-^M {\cal K}^a{}_M(z)+\7Y_+\7Y_-{\cal K}(z)
\nonumber\\
& & +\7Y_+\7A_{a-}^{+*} {\cal K}_{+}^a(z)
+\7A_{a+}^{-*}\7Y_-{\cal K}^a_{-}(z)
+\7A_{a+}^{-*}\7A_{b-}^{+*}{\cal K}^{ab}(z).
\label{Ks}\end{eqnarray}
$\hat \omega$ is denoted analogously, with
a hat-symbol on the functions of the $z$'s contained in it.

Note that the order and positions of the superscripts and subscripts is
important. In particular, in (\ref{Hs})
the functions ${\cal H}^a$ and ${\cal H}_a$ are not related at all, nor 
are the functions ${\cal H}^M$ and 
${\cal H}_M$ (i.e.\ they are not related by
raising or lowering indices). Furthermore ${\cal K}_M{}^a$ and 
${\cal K}^a{}_M$ 
must not be confused.
In the following we will mostly leave out the argument $(z)$.

Using table 3 in section \ref{Result}, one obtains
\begin{eqnarray}
\7s\,\omega &=& 
\left[(\7X_{+}^{M} + \7X_{-}^{M})\partial_{M} + \7Y_{\mu}
\frac{\partial}{\partial\7\Theta_{\mu}} + \7F_{+-}^{a}
\frac{\partial}{\partial\7C^{a}}\right]{\cal F}
\nonumber\\
& &
+\7X_{+-}^{M}({\cal G}^{-}_{M} - {\cal G}^{+}_{M})
+ (\7A^{+*}_{a +-} - \7X_{-}^{M}\partial_{M}D_{a}\cdot){\cal G}^{-a}
- (\7A^{-*}_{a +-} - \7X_{+}^{M}\partial_{M}D_{a}\cdot){\cal G}^{+ a}
\nonumber\\
& &
+ (-)^{\varepsilon_A}y^{A}_{+}\left(\7X_{-}^{M}\partial_{M}
+ \7Y_{-}\frac{\partial}{\partial\7\Theta_{-}}\right){\cal G}^{+}_{A} 
+ (-)^{\varepsilon_A}y^{A}_{-}\left(\7X_{+}^{M}\partial_{M}
+ \7Y_{+}\frac{\partial}{\partial\7\Theta_{+}}\right){\cal G}^{-}_{A}
\nonumber\\
& &
-(\7A^{+*}_{a +-} + \7A^{-*}_{a +-}){\cal H}^{a}
+(\7X_{+}^{M}\7X_{-}^{N}\partial_{M}\partial_{N}D_{a} 
+ \7X_{+-}^{M}\partial_{M}D_{a})\cdot ({\cal H}^{a}_{+} - {\cal H}^{a}_{-})
\nonumber \\[1.5ex]
& &
+ \left[-2G_{MN}\7X_{+-}^{N} +
(H_{KLM} - 2\Gamma_{KLM})\7X_{+}^{K}\7X_{-}^{L} 
+ \7F_{+-}^{a}\partial_{M}D_{a}\cdot\right]{\cal H}^{M}  
\nonumber \\[1.5ex]
& &
+ \left[\7X_{+}^{M}y^{A}_{-}{\cal K}^{a}{}_{A} 
- (-)^{\varepsilon_A}y_{+}^{A}\7X_{-}^{M}{\cal K}_A{}^a\right]
\partial_{M}D_{a} 
\label{s1}
\end{eqnarray}
where $\varepsilon_A$ indicates the Grassmann parity of $y^A_\mu$.
$\7s\, \omega=0$ imposes the following conditions:
\begin{eqnarray}
0 &=&
{\cal H}^{a}-{\cal G}^{-a}= {\cal H}^{a}+ {\cal G}^{+a}
\label{c0}\\
0 &=&
 {\cal H}^{M}\partial_{M}D_{a} + \frac{\partial{\cal F}}{\partial\7C^{a}}
\label{c22}\\
0 &=&
 ({\cal H}^{a}_{+} - {\cal H}^{a}_{-})\partial_{M}D_{a} - 2G_{MN}{\cal H}^{N} 
+ {\cal G}^{-}_{M} - {\cal G}^{+}_{M}
\label{c3}\\
0 &=&
\partial_{M}{\cal F} + {\cal G}^{+a}\partial_{M}D_{a} = 
\partial_{M}{\cal F} - {\cal G}^{-a}\partial_{M}D_{a}
\label{c4a}\\
0 &=&
 \frac{\partial{\cal F}}{\partial\7\Theta_+} = 
\frac{\partial{\cal F}}{\partial\7\Theta_-}
\label{c5}\\
0 &=&
({\cal H}_{+}^{a} - {\cal H}_{-}^{a}) \partial_{M}\partial_{N}D_{a} 
+ (H_{MNK} - 2\Gamma_{MNK}){\cal H}^{K}
\nonumber \\
& & 
- \partial_{N}{\cal G}^{+}_{M} + \partial_{M}{\cal G}^{-}_{N}
+{\cal K}^{a}{}_{N}\partial_{M}D_{a} + {\cal K}_M{}^{a}\partial_{N}D_{a}
\label{c6}\\
0 &=&
 -\frac{\partial{\cal G}^{+}_{M}}{\partial\7\Theta_{-}} + 
\partial_{M}{\cal G}^{-} + {\cal K}^{a}_{-}\partial_{M}D_{a}
\label{c7a}\\
0 &=&
 -\frac{\partial{\cal G}^{-}_{M}}{\partial\7\Theta_{+}} + 
\partial_{M}{\cal G}^{+} - {\cal K}_{+}^{a}\partial_{M}D_{a}
\label{c7b}\\
0 &=&
 \partial_{M}{\cal G}^{-a} + {\cal K}^{b a}\partial_{M}D_{b} = 
\partial_{M}{\cal G}^{+a} - {\cal K}^{ab}\partial_{M}D_{b}
\label{c8}\\
0 &=&
 \frac{\partial{\cal G}^{+}}{\partial\7\Theta_{-}} + 
\frac{\partial{\cal G}^{-}}{\partial\7\Theta_{+}}
\label{c9}\\
0 &=&
\frac{\partial{\cal G}^{-a}}{\partial\7\Theta_{+}} =
\frac{\partial{\cal G}^{+a}}{\partial\7\Theta_{-}}\ .
\label{c100}
\end{eqnarray}
(\ref{c0}), (\ref{c4a}), (\ref{c8}) and
(\ref{c100}) are equivalent to:
\begin{eqnarray}
0&=& \partial_{M}{\cal F} - {\cal H}^{a}\partial_{M}D_{a}
\label{c4}\\
0&=& \partial_{M}{\cal H}^{a} + {\cal K}^{(ab)}\partial_{M}D_{b}
\label{c81}\\
0&=& {\cal K}^{[ab]}\partial_{M}D_{b} = 0\quad \Leftrightarrow\quad 
{\cal K}^{[ab]}=0
\label{c82}\\
0&=& \frac{\partial{\cal H}^{a}}{\partial\7\Theta_{+}}
=\frac{\partial{\cal H}^{a}}{\partial\7\Theta_{-}} 
\label{c10}
\end{eqnarray}
where in (\ref{c82}) we used (\ref{extra}).
(\ref{c3}) determines ${\cal G}^{+}_{M}-{\cal G}^{-}_{M}$, but
not ${\cal G}^{+}_{M}+{\cal G}^{-}_{M}\equiv {\cal P}_M$. Hence, (\ref{c3})
can be replaced by
\begin{equation}
{\cal G}^{\pm}_{M} = \frac{1}{2}{\cal P}_{M} \mp G_{MN}{\cal H}^{N} 
\pm \frac{1}{2}({\cal H}^{a}_{+} - {\cal H}^{a}_{-})\partial_{M}D_{a} \ .
\label{c31}
\end{equation}
Using these expressions, the symmetric and antisymmetric
parts of (\ref{c6}) read
\begin{eqnarray}
& \partial_{(M}[G_{N)K}{\cal H}^K] - \Gamma_{MNK}{\cal H}^{K} + 
\frac{1}{2}{\cal A}^{a}_{(M}\partial_{N)}D_{a} = 0 &
\label{c6a}\\
& \partial_{[M}{\cal P}_{N]} + H_{MNK}{\cal H}^{K} + 
{\cal B}^{a}_{[M}\partial_{N]}D_{a} = 0 &
\label{c6b}
\end{eqnarray}
where
\begin{eqnarray}
{\cal A}^{a}_{M} &=& 
{\cal K}^{a}{}_M + {\cal K}_M{}^{a}- \partial_{M}({\cal H}^{a}_{+} - 
{\cal H}^{a}_{-})
\nonumber\\
{\cal B}^{a}_{M} &=& {\cal K}_{M}{}^{a} - {\cal K}^{a}{}_{M}\ .
\label{AB}
\end{eqnarray}
Using (\ref{c31}), Eqs.\ (\ref{c7a}) and (\ref{c7b}) become
\begin{eqnarray}
\frac{\partial}{\partial\7\Theta_-}\left[ -G_{MN}{\cal H}^N 
+\frac{1}{2}{\cal P}_{M} + \frac{1}{2}({\cal H}^{a}_{+}  
-{\cal H}^{a}_{-})\partial_{M}D_{a}\right] 
= \partial_{M}{\cal G}^{-} + {\cal K}^a_{-}\partial_{M}D_{a}
\nonumber\\
\frac{\partial}{\partial\7\Theta_+}\left[ G_{MN}{\cal H}^N 
+\frac{1}{2}{\cal P}_{M} - \frac{1}{2}({\cal H}^{a}_{+}  
-{\cal H}^{a}_{-})\partial_{M}D_{a}\right] 
= \partial_{M}{\cal G}^{+} 
- {\cal K}_{+}^{a}\partial_{M}D_{a}\ .
\label{c71b}
\end{eqnarray}

We will now further work out the above equations and remove
simultaneously coboundary terms from $\omega$. To this end
we examine how the functions of the $z$'s in $\omega$ change 
under $\omega\rightarrow
\omega-\7s\, \hat \omega$. One finds
\begin{eqnarray}
 {\cal F} & \rightarrow & {\cal F}
\label{r3} \\
 {\cal P}_{M} & \rightarrow & {\cal P}_{M} - 2\partial_{M}\hat {{\cal F}} 
 + (\hat {\cal G}^{-a}-\hat {\cal G}^{+a})\6_M D_a
\label{r5} \\
 {\cal G}^{\pm} & \rightarrow & {\cal G}^{\pm} 
 - \frac{\partial\hat {\cal F}}{\partial \7\Theta_{\pm}}
\label{r8} \\
 {\cal H}^{a} & \rightarrow & {\cal H}^{a} 
\label{r1} \\
 {\cal H}^a_\pm & \rightarrow & 
 {\cal H}^a_\pm \mp \hat {\cal G}^{\mp a}+\hat {\cal H}^a
\label{r1a} \\
 {\cal H}^{M} & \rightarrow & {\cal H}^{M} 
\label{r2} \\
 {\cal H}_M & \rightarrow & {\cal H}_M + \hat {\cal G}^+_M - \hat {\cal G}^-_M
 + 2G_{MN} \hat {\cal H}^N - (\hat {\cal H}^a_+ - \hat {\cal H}^a_-)\6_M D_a
\label{r2a}\\
 {\cal H}_{a} & \rightarrow & 
 {\cal H}_{a} - \hat {\cal H}^{M}\partial_{M}D_{a} 
 - \frac{\partial \hat {\cal F}}{\partial\7C^{a}} 
\label{r10} \\
 {\cal K}_{MN} & \rightarrow & 
 {\cal K}_{MN}+\6_N\hat {\cal G}_M^+ - \6_M\hat {\cal G}_N^- 
 -(H_{MNK} - 2\Gamma_{MNK})\hat {\cal H}^{K}
\nonumber \\
& & 
 -\hat {\cal K}^{a}{}_{N}\partial_{M}D_{a} - 
 \hat {\cal K}_M{}^{a}\partial_{N}D_{a}
 -(\hat {\cal H}^a_+-\hat {\cal H}^a_-)\6_M\6_N D_a
\label{r12} \\
 {\cal K}_{M-}  & \rightarrow & {\cal K}_{M-}
 +\frac{\6 \hat {\cal G}^+_M}{\6 \7\Theta_-}
 -\6_M\hat {\cal G}^--\hat {\cal K}^a_{-}\6_M D_a
\label{r13a}\\
 {\cal K}_{+ M}  & \rightarrow & {\cal K}_{+ M}
 +\frac{\6 \hat {\cal G}^-_M}{\6 \7\Theta_+}
 -\6_M\hat {\cal G}^+
 +\hat {\cal K}_{+}^a\6_M D_a
\label{r13b}\\
 {\cal A}^{a}_{M} & \rightarrow & {\cal A}^{a}_{M} 
 + 2\hat {{\cal K}}^{[ab]}\partial_{M}D_{b}
\label{r6} \\
  {\cal B}^{a}_{M} & \rightarrow & {\cal B}^{a}_{M} 
 - 2\hat {{\cal K}}^{(ab)}\partial_{M}D_{b} 
 - \partial_{M}(\hat {\cal G}^{-a}-\hat {\cal G}^{+a})
\label{r7} \\
 {\cal K} & \rightarrow & {\cal K} 
 - \frac{\partial\hat {\cal G}^{+}}{\partial\7\Theta_{-}}
- \frac{\partial\hat {\cal G}^{-}}{\partial\7\Theta_{+}}
\label{r11} \\
 {\cal K}_{+}^a & \rightarrow & {\cal K}_{+}^a
 -\frac{\partial\hat {\cal G}^{-a}}{\partial\7\Theta_{+}}
\label{r9a} \\
 {\cal K}^a_{-} & \rightarrow & {\cal K}^a_{-}
 -\frac{\partial\hat {\cal G}^{+a}}{\partial\7\Theta_-}
\label{r9b} \\
 {\cal K}^{(ab)} & \rightarrow & {\cal K}^{(ab)}
\label{r4} 
\end{eqnarray}
where we have used already (\ref{c31}) and (\ref{AB}),
i.e.\ we have switched from the functions
${\cal G}_M^+ + {\cal G}_M^-$, ${\cal K}^{a}{}_M$ and ${\cal K}_M{}^{a}$ to
${\cal P}_M$, ${\cal A}^{a}_{M}$, ${\cal B}^{a}_{M}$. 
(\ref{r1a}) and (\ref{r2a}) show
that we can always remove ${\cal H}^a_+$, ${\cal H}^a_-$ and ${\cal H}_M$
completely from $\omega$ by choosing
$\hat {\cal G}^{\mp a}=\pm ({\cal H}^a_\pm +\hat {\cal H}^a)$ and
$\hat {\cal G}^\mp_M = \frac 12\hat {\cal P}_M \pm \frac 12{\cal H}_M  
 \pm G_{MN} \hat {\cal H}^N \mp \frac 12(\hat {\cal H}^a_+ - 
 \hat {\cal H}^a_-)\6_M D_a$
where $\hat {\cal P}_M$ is still arbitrary.
In the following we will assume that ${\cal H}^a_+$, ${\cal H}^a_-$ and 
${\cal H}_M$
have been removed in this way.
Note that this choice of
$\hat {\cal G}^{\mp a}$ and $\hat {\cal G}^\mp_M$ shifts also ${\cal P}_M$, 
${\cal K}_{MN}$,
${\cal K}_{M-}$, ${\cal K}_{+ M}$, ${\cal B}_M^a$,
${\cal K}_{+}^a$ and ${\cal K}^a_{-}$. For simplicity
we denote the shifted functions by the same symbols as the original ones. 
Then we can still work with the above formulae,
setting ${\cal H}^a_+$, ${\cal H}^a_-$ and ${\cal H}_M$ to zero everywhere.
Hence, without loss of generality we can assume
\begin{eqnarray}
 {\cal H}^a_\pm &=& {\cal H}_M\, =\, 0
\label{choice2}\\
 \hat {\cal G}^{\mp a} &=& \pm \hat {\cal H}^a
\label{choice3}\\ 
 \hat {\cal G}^\mp_M &=& \frac 12\hat {\cal P}_M
 \pm G_{MN} \hat {\cal H}^N \mp 
 \frac 12 (\hat {\cal H}^a_+ - \hat {\cal H}^a_-)\6_M D_a
\label{choice4}
\end{eqnarray}
where (\ref{choice3}) and (\ref{choice4}) guarantee 
that (\ref{choice2}) is preserved
by the remaining redefinitions (\ref{r3})--(\ref{r4}).

In addition to (\ref{choice2}), we can remove certain parts of
other functions occurring in $\omega$ through the above redefinitions.
To this end we use the superfield notation as in (\ref{superfield}),
leaving out the arguments $(X,\7C)$
of the component functions.

(\ref{c5}) and (\ref{c10}) require evidently
\begin{eqnarray}
&&
f^\pm=\5f=h^{a\pm}=\5h^a=0\quad \Leftrightarrow \quad
{\cal F}=f\ ,\quad {\cal H}^a=h^a
\label{choice0}
\end{eqnarray}
(\ref{c9}) requires $g^{+-}=-g^{-+}$
and $\5g^+=\5g^-=0$. 
Furthermore, (\ref{r8})
gives $g^\pm\rightarrow g^\pm
-\hat f^\pm$ and $g^{+-}\rightarrow g^{+-}
-\overline{\hat f}$. Hence, we can always remove
$g^+$, $g^-$, $g^{+-}$ and
$g^{-+}$ from $\omega$ by choosing $\hat f^\pm$
and $\overline{\hat f}$ appropriately. By the same reasoning
that has led to (\ref{choice2})--(\ref{choice4}) we can therefore
assume, without loss of generality,
\begin{eqnarray}
g^\pm=g^{+-}
=g^{-+}=\5g^\pm=0
&\Leftrightarrow &
{\cal G}^{\pm} = \7\Theta_\pm\, g^{\pm\pm}
\label{choice5}\\
\hat f^\pm=\overline{\hat f}=0
&\Leftrightarrow &
\hat {\cal F} = \hat f
\label{choice6}
\end{eqnarray}
Analogously we conclude from
(\ref{r11})--(\ref{r9b}),
taking (\ref{choice3}) into account,
that without loss of generality we can assume
\begin{eqnarray}
k=k^\pm=0
& \Leftrightarrow & 
{\cal K}=\7\Theta_+\7\Theta_-\5k
\label{choice7}
\\
\bar{\hat g}^\pm=0,\ 
\hat g^{+-}=-\hat g^{-+}\equiv \hat g
&\Leftrightarrow &
\hat {\cal G}^\pm=\hat g^\pm
\pm\7\Theta_\mp\,\hat g+\7\Theta_\pm\,\hat g^{\pm\pm}
\label{choice10}
\end{eqnarray}
and
\begin{eqnarray}
k_{+}^a=k^a_{-}
=\hat h^{a\pm}=\bar{\hat h}{}^a=0,\quad
k^{a+}_{-}=-k_{+}^{a-}\equiv k^a
\nonumber\\
\Leftrightarrow \ \left\{
\begin{array}{l} 
{\cal K}_{+}^a=-\7\Theta_-k^a+\7\Theta_+k_{+}^{a+}
+\7\Theta_+\7\Theta_-\5k_{+}^a
\\
{\cal K}^a_{-}=\7\Theta_+k^a+\7\Theta_-k^{a-}_{-}
+\7\Theta_+\7\Theta_-\5k^a_{-}
\\
\hat{\cal H}^a=\hat h^a
\end{array}
\right.
\label{choice8}
\end{eqnarray}
Taking into account (\ref{choice4}) and (\ref{choice10}), Eqs.
(\ref{r13a}) and (\ref{r13b}) yield analogously
\begin{eqnarray}
& &
k_{M-}=k_{+ M}=0,\quad 
k_{M-}^+=k_{+ M}^-\equiv k_M
\nonumber\\
& &\Leftrightarrow\ \left\{ 
\begin{array}{l}
{\cal K}_{M-}=\7\Theta_+k_M+\7\Theta_-k_{M-}^-
+\7\Theta_+\7\Theta_-\5k_{M-}
\\
{\cal K}_{+ M}=\7\Theta_-k_M+\7\Theta_+k_{+ M}^+
+\7\Theta_+\7\Theta_-\5k_{+ M}\ ,
\end{array}
\right.
\label{choice13} 
\end{eqnarray}
as well as Eqs. (\ref{choice14a})--(\ref{choice14c}) up to an extra 
contribution
$2\6_M\hat g$ to $\overline{\hat p}_M$ which can be neglected
in (\ref{choice14c}) as it provides an $\7s$-exact contribution
to $\hat\omega$ (furthermore we have used the notation
$\hat {\cal H}_+^a-\hat {\cal H}_-^a=2\hat {\cal Z}^a$ 
in section \ref{Result}).

Next we analyse Eqs.\ (\ref{c71b}), using
(\ref{choice2}), (\ref{choice5}) and
(\ref{choice8}).
The $\7\Theta$ independent components of Eqs.\ (\ref{c71b}) yield
\begin{equation}
p_M^\pm=\mp 2G_{MN}h^{N\pm}\ ,
\label{choice15}
\end{equation}
the terms linear in the $\7\Theta$'s 
give
\begin{eqnarray}
\5p_M &=& 0
\label{choice16}\\
G_{MN}\5h^N &=& k^a\6_M D_a
\label{choice17}\\
\6_M g^{\pm\pm} &=& \pm k_{\pm}^{a\pm}\6_M D_a\ ,
\label{choice18}
\end{eqnarray}
and the $\7\Theta_+\7\Theta_-$ components
yield
\begin{equation}
\5k^a_{-}\6_M D_a=\5k_{+}^a\6_M D_a=0
\quad\Leftrightarrow\quad 
\5k^a_{-}=\5k_{+}^a=0
\label{choice19}
\end{equation}
where we used (\ref{extra}).

Inserting (\ref{choice0}) in (\ref{c81}) we get in particular
\begin{equation}
k^{(ab)\pm}\6_M D_a=\5k^{(ab)}\6_M D_a=0
\quad \Leftrightarrow\quad 
k^{(ab)\pm}=\5k^{(ab)}=0
\label{zusatz}\end{equation}
where we used again (\ref{extra}). 

Finally we examine the $\7\Theta_+ \7\Theta_-$ components of
(\ref{c6a}), (\ref{c6b}) and (\ref{c22}). Using
(\ref{choice0}), (\ref{choice16}) and (\ref{choice17}) one gets
\begin{eqnarray}
& \partial_{(M}[k^a \6_{N)} D_a] - \Gamma_{MNK}\5h^{K} 
+\frac 12\, \5a^{a}_{(M}\partial_{N)}D_{a} = 0 &
\nonumber\\
& H_{MNK}\5h^{K} + \5b^{a}_{[M}\partial_{N]}D_{a} = 0 &
\nonumber\\
& \5h^M \6_M D_a = 0,\quad G_{MN}\5h^N = k^a \6_M D_a\ . &
\label{cond8}
\end{eqnarray}
Due to (\ref{extra1a})--(\ref{extra2}) we conclude from
(\ref{cond8}), using (\ref{r6}) and (\ref{r7}), that we
can assume without loss of generality
\begin{equation}
\5h^M=k^a=\5a_M^a=\5b_M^a=\overline{\hat k}{}^{ab}=0.
\end{equation} 
Altogether this provides the results given in section \ref{Result}.

\mysection{Nonabelian deformations}\label{appC}

In this appendix we describe some obvious deformations arising
from (\ref{W06}) and show how nonabelian Born-Infeld type actions
emerge from them in special cases.

Assume we choose
a Lagrangian (\ref{L}) which is invariant under the global action
of a compact Lie group represented linearly on the $X^M$ and $A_\mu^a$,
such that the representation of the $A_\mu^a$ is the adjoint one
(i.e., the indices $a$ of the abelian gauge fields label
simultaneously the generators of the Lie group). That is to say,
assume
\begin{eqnarray*}
\delta_a X^M=-T_{a N}^M X^N,\quad
\delta_a A_\mu^b=-f_{ac}{}^b A_\mu^c\quad\Rightarrow\quad \delta_a L=0
\end{eqnarray*}
where $\delta_a$ denote the generators of the Lie group,
$f_{bc}{}^a$ are the structure constants of the 
corresponding Lie algebra, and
$\{T_a\}$ is the matrix representation of that Lie algebra
on the $X^M$,
\[ \mbox{\bf [}\, T_a\, ,\, T_b \, \mbox{\bf ]}
=f_{ab}{}^c\, T_c\ .\]
This requires of course the functions 
$G_{MN}(X)$, $B_{MN}(X)$ and $D_a(X)$ to
transform under ${\cal G}$ according to the representation 
indicated by their indices. Then it is clear that a nonabelian
deformation of the Lagrangian is obtained just by the replacements
\begin{eqnarray*}
\6_\mu A_\nu^a- \6_\nu A_\mu^a & \rightarrow & 
\6_\mu A_\nu^a- \6_\nu A_\mu^a+g f_{bc}{}^a A_\mu^b A_\nu^c
\\
\6_\mu X^M & \rightarrow & D_\mu X^M=
\6_\mu X^M+g A_\mu^a T_{a N}^M X^N
\end{eqnarray*}
where $g$ is a constant deformation parameter.
The deformed Lagrangian is invariant under
deformed gauge transformations
\begin{eqnarray*}
\delta_\varepsilon X^M=-g \varepsilon^a T_{a N}^M X^N,\quad
\delta_\varepsilon A_\mu^a=
\6_\mu \varepsilon^a+g f_{bc}{}^a \varepsilon^c A_\mu^b\ .
\end{eqnarray*}
To first order in $g$, these deformations are indeed given
by (\ref{W06}) with
\begin{eqnarray*}
h_{bc}^a = f_{bc}{}^a,\ 
h^M_a = -T_{a N}^M X^N,\ 
p'_{Ma}=b_{b M}^a = a_{b M}^a = k_{cd}^{(ab)} = 0.
&
\end{eqnarray*}
This holds as ${\cal L}_a G_{MN}={\cal L}_a B_{MN}=0$ are 
implied by the assumption that $G_{MN}(X)$ and $B_{MN}(X)$ transform
under ${\cal G}$ as indicated by their indices.

Nonabelian Born-Infeld type actions are obtained from the
above formulae as follows. We split the $X^M$ into two
subsets,
\[
\{X^M\}=\{x^m,\varphi^a\},
\]
where the $\varphi^a$ transform according to the adjoint
representation of some compact Lie group.
For simplicity we assume that the $x^m$ transform according to
the trivial representation, i.e.
\[ T_{ab}^c=-f_{ab}{}^c,\quad T_{a m}^n=0. \]
We now choose a faithful matrix representation
$\{\mbox{{\bf t}}_a\}$ of the Lie algebra and introduce the matrix
\[
\mbox{{\boldmath $\varphi$}}=\varphi^a\, \mbox{{\bf t}}_a\ .
\]
Next we construct a Lagrangian (\ref{L}) with
\begin{eqnarray*}
& 
G_{mn}(X) = g_{mn}(x)\, \mbox{Tr} (\mbox{{\bf 1}}
-\mbox{{\boldmath $\varphi$}}^2)^{-1/2},\quad
G_{ma}=G_{ab}=B_{MN}=0
&
\\
&
D_a(X)=-\mbox{Tr} [\mbox{{\bf t}}_a\, \mbox{{\boldmath $\varphi$}}\, 
(\mbox{{\bf 1}}-\mbox{{\boldmath $\varphi$}}^2)^{-1/2}]
&
\end{eqnarray*}
where $(\mbox{{\bf 1}}-\mbox{{\boldmath $\varphi$}}^2)^{-1/2}$ 
is understood as a power series
in $\mbox{{\boldmath $\varphi$}}$. 
A nonabelian deformation of the corresponding Lagrangian
is now given by
\[
L^{nonabel.}_0=\frac 12\, 
\mbox{Tr}\left\{(\mbox{{\bf 1}}-\mbox{{\boldmath $\varphi$}}^2)^{-1/2}
(\mbox{{\bf 1}}\sqrt\gamma\, \gamma^{\mu\nu} g_{mn}(x)\,
\6_\mu x^m\cdot \6_\nu x^n
-\epsilon^{\mu\nu}\mbox{{\boldmath $\varphi$}} 
\mbox{{\bf F}}_{\mu\nu})\right\}
\]
where
\[
\mbox{{\bf F}}_{\mu\nu}=
(\6_\mu A_\nu^a- \6_\nu A_\mu^a+g f_{bc}{}^a A_\mu^b A_\nu^c)\, 
\mbox{{\bf t}}_a\ .
\]
Elimination of the auxiliary fields
$\gamma_{\mu\nu}$ and $\varphi^a$ yields (\ref{nonabelian}).
Note that this example can be easily generalized by allowing the
$x^m$ to transform linearly under the Lie group, or nontrivial
$B_{mn}$ in the way outlined above. Note also that an 
ordering ambiguity of matrices under the above traces does not
arise.


\begin{thebibliography}{99}

\bibitem{joe}
See J. Polchinski, Tasi Lectures on D-branes, hep-th/9611050
and refs. therein.

\bibitem{paul}
P. K. Townsend, {\em Phys. Lett.} {\bf 409B} (1997) 131.

\bibitem{cederwall}
M. Cederwall and P. K. Townsend, JHEP 09 (1997) 003
(hep-th/9709002);\\
M. Cederwall and A. Westerberg, World volume fields,
SL(2,Z) and duality: the type IIB three-brane,
hep-th/9710007.

\bibitem{stevequim}
J. Gomis and S. Weinberg,
{\em Nucl. Phys.} {\bf B469} (1996) 473.

\bibitem{bbh1}
G. Barnich, F. Brandt and M. Henneaux,
{\em Commun. Math. Phys.} {\bf 174} (1995) 57.

\bibitem{bh}
G. Barnich and M. Henneaux, 
{\em Phys. Lett.} {\bf 311B} (1993) 123.

\bibitem{anos}
This was first observed in the pioneering works
C. Becchi, A. Rouet and R. Stora, 
{\em Phys. Lett.} {\bf 52B} (1974) 344;
{\em Commun. Math. Phys.} {\bf 42} (1975) 127; 
{\em Ann. Phys.} {\bf 98} (1976) 287.\\
See also \cite{henteit,report} and refs.\ therein.

\bibitem{bv} 
I. A. Batalin and G. A. Vilkovisky, {\em Phys. Lett.}
{\bf 102B} (1981) 27; {\em Phys. Rev.} {\bf D28} (1983) 2567
(E: {\bf D30} (1984) 508).

\bibitem{henteit}
M. Henneaux and C. Teitelboim, {\em Quantization of
Gauge Systems} (Princeton University Press, Princeton, 1992).

\bibitem{report}
J. Gomis, J. Par\'{\i}s and S. Samuel,
{\em Phys. Rep.} {\bf 259} (1995) 1.

\bibitem{KT}
J. M. L. Fisch and M. Henneaux, {\em Commun. Math. Phys.}
{\bf 128} (1990) 627;\\
M. Henneaux, 
{\em Nucl. Phys. B (Proc. Suppl.)} {\bf 18A} (1990) 47.\\
See also \cite{henteit}.

\bibitem{ten}
F. Brandt, {\em Commun. Math. Phys.} {\bf 190} (1997) 459.

\bibitem{paper1}
F. Brandt, W. Troost and A. Van Proeyen,
{\em Nucl. Phys.} {\bf B464} (1996) 353.

\bibitem{letter}
F. Brandt, J. Gomis and J. Sim\'on,
{\em Phys. Lett.} {\bf 419B} (1998) 148.

\bibitem{bhw}
F. Brandt, M. Henneaux and A. Wilch, {\em Nucl. Phys. (PM)}
{\bf B510} (1998) 640.

\bibitem{Hull}
B. de Wit and P. van Nieuwenhuizen, {\em Nucl. Phys.} 
{\bf B312} (1989) 58;\\
C. M. Hull and B. Spence, {\em Phys. Lett.} {\bf 232B} (1989) 204.

\bibitem{nonabBI}
A. A. Tseytlin, {\em Nucl. Phys.} {\bf B469} (1996) 51.

\bibitem{paper2}
F. Brandt, W. Troost and A. Van Proeyen,
{\em Phys. Lett.} {\bf 374B} (1996) 31.

\bibitem{comgrav}
F. Brandt, N. Dragon and M. Kreuzer,
{\em Phys. Lett.} {\bf 231B} (1989) 263;
{\em Nucl. Phys.} {\bf B332} (1990) 224;
{\em Nucl. Phys.} {\bf B340} (1990) 187.

\bibitem{2U1s}
F. Brandt, {\em Phys. Lett.} {\bf 320B} (1994) 57;\\
G. Barnich and M. Henneaux,
{\em Phys. Rev. Lett.} {\bf 72} (1994) 1588;\\
G. Barnich, F. Brandt and M. Henneaux,
{\em Commun. Math. Phys.} {\bf 174} (1995) 93;
{\em Nucl.\ Phys.} {\bf B455} (1995) 357.

\end{thebibliography}
\end{document}